\documentclass[journal]{IEEEtran}
\ifCLASSINFOpdf
\else
\fi
\usepackage{cite}
\usepackage{amsmath,amssymb,amsfonts}
\usepackage{algorithmic}
\usepackage{graphicx}
\usepackage{textcomp}
\usepackage{xcolor}
\def\BibTeX{{\rm B\kern-.05em{\sc i\kern-.025em b}\kern-.08em
		T\kern-.1667em\lower.7ex\hbox{E}\kern-.125emX}}

\usepackage{mathtools}

\DeclarePairedDelimiter\floor{\lfloor}{\rfloor}
\usepackage{steinmetz}
\usepackage{authblk}
\newcommand{\minus}{\scalebox{0.75}[1.0]{$-$}}

\usepackage{blindtext}
\usepackage{caption}
\usepackage{subcaption}

\begin{document}
%
\title{Rate-Splitting Multiple Access to Mitigate the Curse of Mobility in (Massive) MIMO Networks}
%
%
%

\author{Onur~Dizdar,~\IEEEmembership{Member,~IEEE,}
        Yijie~Mao,~\IEEEmembership{Member,~IEEE,}
        and~Bruno Clerckx,~\IEEEmembership{Senior~Member,~IEEE}
\thanks{The authors are with the Department of Electrical and Electronic Engineering, Imperial College London, London, UK. \mbox{e-mail: (\{o.dizdar,y.mao16,b.clerckx\}@imperial.ac.uk)}.\\This research was partially funded by the U.K. Engineering and Physical Sciences Research Council (EPSRC) under grant EP/R511547/1.}
}

\maketitle

\begin{abstract}
	Rate-Splitting Multiple Access (RSMA) is a robust multiple access scheme for downlink multi-antenna wireless networks. RSMA relies on multi-antenna Rate-Splitting (RS) at the transmitter and Successive Interference Cancellation (SIC) at the receivers. In this work, we study the performance of RSMA under the important setup of imperfect Channel State Information at the Transmitter (CSIT) originating from user mobility and latency/delay (between CSI acquisition and data transmission) in the network. We derive a lower bound on the ergodic sum-rate of RSMA for an arbitrary number of transmit antennas, number of users, user speed and transmit power. Then, we study the power allocation between common and private streams and obtain a closed-form solution for optimal power allocation that maximizes the obtained lower bound. The proposed power allocation greatly reduces precoder design complexity for RSMA. By Link-Level Simulations (LLS), we demonstrate that RSMA with the proposed power allocation is robust to the degrading effects of user mobility and has significantly higher performance compared to conventional multi-user (massive) Multiple-Input Multiple-Output (MIMO) strategies. The work has important practical significance as results demonstrate that, in contrast to conventional multi-user (massive) MIMO whose performance collapse under mobility, RSMA can maintain reliable multi-user connectivity in mobile deployments.    
\end{abstract}

\begin{IEEEkeywords}
Rate-splitting, multi-antenna broadcast channel, outdated CSIT, ergodic sum-rate.
\end{IEEEkeywords}

%
\IEEEpeerreviewmaketitle

\section{Introduction}
\label{sec:introduction}
\IEEEPARstart{A}{n} essential requirement for the next generation communication systems is to meet the demand for exponentially increasing data rate and accessibility. A key technology expected to meet such requirement is small-scale and large-scale (massive) Multi-User Multiple-Input Multiple-Output (MU-MIMO). Aside having been studied in a vast amount of literature, MU-MIMO concept has also been used in real life wireless communications standards, with 4G and 5G being among the most notable ones. 
The performance achieved by beamforming and interference management methods in MU-MIMO heavily relies on the Channel State Information (CSI) accuracy. However, acquiring accurate CSI at the Transmitter (CSIT) is a major challenge in real life scenarios. CSIT is commonly subject to errors, which is known as imperfect CSIT.
Imperfect CSIT is a common problem in wireless communication networks and causes severe multi-user interference in multi-antenna Broadcast Channel (BC).

A major cause of imperfect CSIT in modern wireless communication networks is the CSI acquisition/feedback latency from mobile users, which results in outdated CSIT due to the variations in the wireless medium under mobility. Indeed, one of the earliest problems reported in field trials for MU-MIMO communications in 5G is outdated CSIT due to mobility. The processing delay in the system easily becomes larger than the coherence time of the channel even
when the users are moving with a speed of $30$km/h, making the CSIT inaccurate for multi-user beamforming \cite{wang_2017}. 
Moreover, mobility scenarios are expected to become increasingly important in next generation wireless communication systems.
In addition to the broadband data services, use cases such as Vehicle-to-Everything (V2X), very high speed trains and transportation, and very low earth orbit satellite systems, are especially prone to the disruptive effects of mobility. Therefore, maintaining high-rate multi-user connectivity and communications in mobile networks becomes a crucial and fundamental problem in future communication systems \cite{tariq_2019, strinati_2019, dizdar_2020_2}.
Therefore, it is of paramount importance to seek strategies that can mitigate the detrimental effects induced by user mobility and latency in multi-user multi-antenna communications.

The most common method of achieving MU-MIMO communications is to use multi-user linear precoding techniques to separate users' streams at the transmitter. At the receivers, multi-user interference is treated as noise during signal detection. Such strategy is often named as Space Division Multiple Access (SDMA). 
A major drawback of SDMA is the performance degradation due to the abovementioned problem of imperfect CSIT. 
Studies investigating both theoretical (e.g., Degrees-of-Freedom (DoF) \cite{clerckx_2016}) and practical (e.g., throughput \cite{wang_2017}) metrics demonstrate the dependence of SDMA performance on the CSIT quality. Therefore, maintaining multi-user connectivity and performance improvements under mobility conditions remain a crucial problem for SDMA-assisted Multi-User (Massive) MIMO. 

A method to tackle the issues of latency and mobility in SDMA-assisted Multi-User (Massive) MIMO is to perform switching between Single-User (SU) and Multi-User (MU) transmission modes \cite{zhang_2009, zhang_2011}. Such an approach restricts transmission to a small number of users (even to the SU case in the extreme case) so as to reduce the interference level and prevent a sum-rate loss in the network. Consequently, MU communications cannot be maintained unless interference is reduced to certain levels. 
Another strategy is to play at the channel acquisition level and conduct channel prediction and reconstruction at the transmitter \cite{truong_2013, papazafeiropoulos_2015, kong_2015, kashyap_2017, yin_2019}. However, the complexity of channel prediction algorithms becomes quickly overwhelming with an increasing number of transmit and receive antennas and subcarriers in an Orthogonal Frequency Division Multiplexing (OFDM) MU-MIMO systems.

In this work, we depart from conventional precoding techniques and take a different approach that is inherently robust to imperfect CSIT and multi-user interference to tackle the mobility problem. 
We consider Rate-Splitting Multiple Access (RSMA), which is a multiple access technique for multi-antenna Broadcast Channel (BC) that relies on linearly precoded Rate-Splitting (RS) at the transmitter and Successive Interference Cancellation (SIC) at the receivers \cite{clerckx_2016, mao_2018}. RSMA is based on splitting the user messages into common and private parts at the transmitter. The common parts of the split messages of each user are combined and encoded into a single common message to be decoded by all users. The single common message and the private parts of the user messages are encoded separately into common and private streams, precoded and superposed in a non-orthogonal manner. Thanks to the message split and the presence of common and private streams, RSMA manages interference by partially decoding the interference and treating the remaining interference as noise by means of SIC receivers. 

Originally introduced for the Single-Input Single-Output (SISO) Interference Channel (IC) \cite{carleial_1978, han_1981}, RS has been adapted to multi-antenna BC due to recent information theoretic studies showing that RS achieves the entire DoF region of $K$-user Multiple-Input Single-Output (MISO) BC with partial CSIT \cite{piovano_2017,joudeh_2016_2}, and outperforms Dirty Paper Coding (DPC) \cite{mao_2020}. 
The performance gain of RSMA comes from its ability to manage the interference in a flexible manner (namely bridging the extremes of treating multi-user interference as noise, as in SDMA, and fully decoding the multi-user interference, as in Non-Orthogonal Multiple Access (NOMA)) by splitting the messages of the users and performing SIC at the receivers.
Being a multiple access scheme designed for BC with multiple transmit (and receive) antennas, RSMA is naturally a promising candidate for operating under interference in multi-antenna networks. 
Indeed, numerous studies have demonstrated that RSMA bridges and outperforms existing multiple access schemes, i.e., SDMA, NOMA, Orthogonal Multiple Access (OMA) and multicasting \cite{joudeh_2016_2,clerckx_2019,mao_2018}. In its simplest form, 1-layer RSMA relies on a single common stream and requires a single SIC at each receiver \cite{mao_2018}.

In the context of RSMA use in practical systems, 1-layer RSMA with low-complexity precoders for the common and private streams is considered. A problem arising in such system is how to allocate the total transmit power between the common and private streams in an optimal manner for an improved sum-rate performance under imperfect CSIT. 

\subsection{Related Work}
An analysis of sum-rate for 1-layer RSMA for the two-user MISO BC is studied in \cite{clerckx_2020} with ZF precoders for the private streams and perfect CSIT. The obtained sum-rate expression is used to derive a power allocation method to distribute power between the common and private streams for the considered scenario. 

The authors investigate power allocation between the common and private streams of 1-layer RSMA for the two-user MISO BC in \cite{hao_2015} under channel quantization error and ZF precoders for the private streams. The channel estimation error is bounded (i.e, isotropically distributed) and the interference from the private stream of the unintended user is neglected, which is reasonable for a scenario with two users and a low level of channel quantization error. Equal power allocation is considered among the private streams of the two users. The obtained power allocation method is optimal in the sense that it minimizes the ergodic sum-rate difference of SDMA with perfect CSIT and RSMA with imperfect CSIT. 

Another work on the power allocation between the common and private streams is \cite{dai_2016}, which considers a massive MIMO setup with imperfect CSIT and unbounded channel estimation error (i.e, Complex Gaussian distributed). The sum-rate analysis and obtained closed-form power allocation algorithm are based on the asymptotic case of a large number of transmit antennas and number of users. 

Precoder design and power allocation based on second-order channel statistics are investigated for RSMA with hybrid precoding for Millimeter Wave (mmWave) in \cite{dai_2017}. 
In the low Signal to Noise (SNR) regime, the proposed power allocation assigns all transmit power to the private streams. On the contrary, in the high SNR regime, only a portion of the total transmit power is assigned to the private streams such that the private streams are not interference limited, {\sl i.e.}, the noise power remains larger or equal to the interference left between the private streams.

The studies in the literature so far have investigated specific cases in terms of the number of transmit antennas and/or users, and/or resort to optimization frameworks for RSMA \cite{mao_2018,joudeh_2016_2} that do not lend themselves to practical implementations. Therefore, obtaining a power allocation algorithm under imperfect CSIT for an arbitrary number of transmit antennas and users (and not necessarily infinitely many) still remains an open problem. 
Furthermore, there is a lack of consideration for the imperfections in the channel estimate quality originating from user mobility and latency in the network. The severity of CSIT quality degradation under user mobility causes non-negligible interference terms in the rate expressions, making it more difficult to analytically design practical power allocation algorithms.

\subsection{Contributions}
The contributions of this paper can be listed as follows:
\begin{itemize}
	\item RSMA has been shown to outperform existing multiple access schemes under imperfect CSIT resulting from imperfect channel estimation and quantized feedback. In this work, we design RSMA specifically for CSIT imperfections caused by user mobility in Multi-User (Massive) MIMO. This is the first paper that shows that RSMA can be used to maintain reliable multi-user connectivity and mitigate the curse of mobility in Multi-User MIMO networks. 
	\item We propose a system model where a multi-antenna transmitter employing RSMA serves multiple single-antenna users with mobility. We consider an arbitrary number of transmit antennas, number of users, total transmit power and user speed in the system. We consider low-complexity transmit precoders for RSMA for the common and private streams. The splitting of the user rates is controlled by a power allocation coefficient that distributes the total transmit power between common and private precoders.  
	\item Taking user mobility into consideration, we derive a lower bound for the approximated ergodic sum-rate of 1-layer RSMA for an arbitrary number of transmit antennas, number of users, amount of total transmit power and user mobility in a tractable form. 
	The analytical expressions show that the obtained lower bound is also a function of the power allocation coefficient that distributes the total transmit power between common and private precoders.   
	We demonstrate by simulations that the obtained lower bound captures the behaviour of the ergodic sum-rate for different sets of system parameters and different values of the power allocation coefficient. 
	\item Using the derived lower bound, we obtain a closed-form solution for the power allocation coefficient that distributes the total transmit power between the common stream and private streams and propose a power allocation algorithm based on the obtained closed-form solution. 
	The obtained closed-form solution is optimal in the sense that it maximizes the lower bound for the approximated ergodic sum-rate. 
	The proposed power allocation method is valid for any number of transmit antennas, number of users, total transmit power and user speed. 
	\item We perform simulations to analyse the performance of RSMA with the proposed power allocation method. We use Monte-Carlo and Link-Level Simulations (LLS) based on finite constellations and finite length polar codes to investigate the sum-rate and throughput performance. Furthermore, we compare the performance of RSMA and SDMA to highlight the significant sum-rate gain offered by RSMA.
	The results demonstrate that, in contrast to SDMA and conventional Multi-User (massive) MIMO whose performance collapse under mobility, RSMA can maintain multi-user connectivity in mobile multi-antenna systems. The results also show that the benefits are visible in a wide range of deployments, from small MIMO regime to large-scale (massive) MIMO regime, with a small and a large number of users.
\end{itemize}

The rest of the paper is organized is as follows. Section~\ref{sec:system} gives the system model. We derive the lower bound for the ergodic sum-rate of 1-layer RSMA with ZF precoding in Section~\ref{sec:lowerboundergodic}. We obtain a closed-form expression and perform an asymptotic analysis for the power allocation algorithm for RSMA in Section~\ref{sec:closedform}. Section~\ref{sec:simulation} gives the simulation results on the performance of RSMA with the proposed power allocation algorithm. Section~\ref{sec:conclusion} concludes the paper.

\textit{Notation:} Vectors are denoted by bold lowercase letters. 
The operations $|.|$ and $||.||$ denote the cardinality of a set or absolute value of a scalar and $l_{2}$-norm of a vector, respectively. 
$\mathbf{a}^{T}$ and $\mathbf{a}^{H}$ denote the transpose and Hermitian transpose of a vector $\mathbf{a}$, respectively. 
$\mathcal{CN}(0,\sigma^{2})$ denotes the Circularly Symmetric Complex Gaussian distribution with zero mean and variance $\sigma^{2}$. $\mathbf{I}$ denotes the identity matrix. $\left \lfloor{.}\right \rceil $ denotes the round operation. $\mathrm{Gamma}(D,\theta)$ represents the Gamma distribution with the probability density function $f(x)=\frac{1}{\Gamma(D)\theta^{D}}x^{D-1}e^{-\frac{x}{\theta}}$. For any complex $x$ with a positive real part, $\Gamma(x)=\int_{0}^{\infty}t^{x-1}e^{-t}dt$ is the gamma function and $\Gamma^{\prime}(x)$ is the derivative of $\Gamma(x)$ with respect to $x$. 

\section{System Model}
\label{sec:system}
\begin{figure*}[t!]
	\centerline{\includegraphics[width=4.5in,height=4.5in,keepaspectratio]{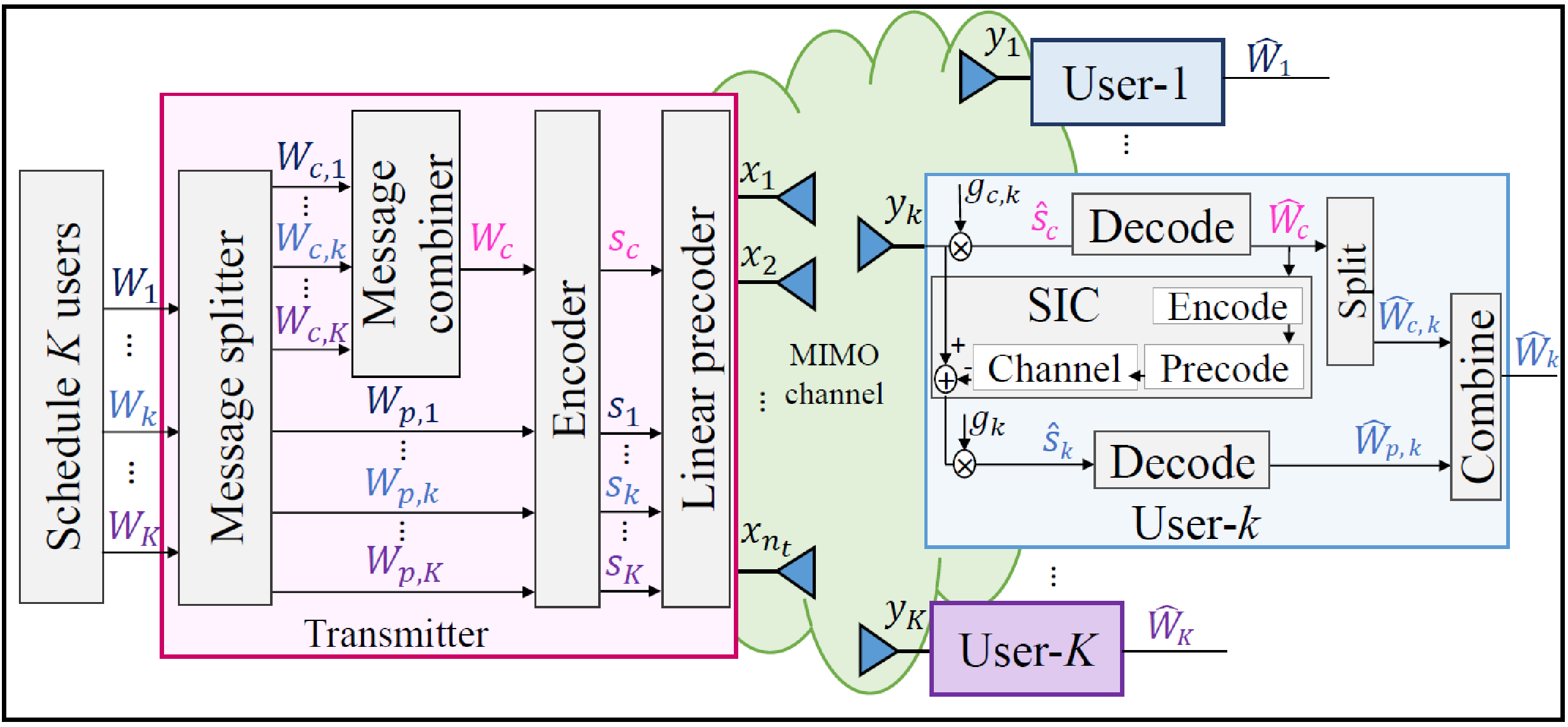}}
	\vspace{-0.2cm}
	\caption{Transmission model of $K$-user 1-layer RSMA.}
	\label{fig:system}
	\vspace{-0.4cm}
\end{figure*}
We consider the MISO BC setting consisting of one transmitter with $n_{t}$ transmit antennas and $K$ single-antenna users indexed by $\mathcal{K}=\{1,\ldots,K\}$, with $n_{t} \geq K$. We consider RSMA for multi-user transmission. Fig.~\ref{fig:system} shows the transmission model of 1-layer RSMA with $K$ users. 
RSMA relies on splitting the user messages at the transmitter and partially decoding the interference and partially treating the remaining interference as noise at the receivers.

The message $W_{k}$, which is intended for user-$k$, is split into one common part $W_{c,k}$ and one private part $W_{p,k}$, $\forall k\in\mathcal{K}$. The split messages are assumed to be independent. The common parts of the messages, $W_{c,k}$, $k\in \mathcal{K}$, are combined into the single common message $W_{c}$. The common message $W_{c}$ and the $K$ private messages $W_{p,k}, k\in\mathcal{K}$ are independently encoded into streams $s_{c}$, $s_{k}$, $k\in\mathcal{K}$, respectively. Linear precoding is applied to all streams. The transmit signal is expressed as
\begin{align}
	\mathbf{x}=\sqrt{P(1-t)}\mathbf{p}_{c}s_{c}+\sqrt{\frac{Pt}{K}}\sum_{k \in \mathcal{K}}\mathbf{p}_{k}s_{k}.
	\label{eqn:1}
\end{align}
The precoders $\mathbf{p}_{c}, \mathbf{p}_{k}\in\mathbb{C}^{n_t}$ satisfy $||\mathbf{p}_{c}||^{2}=1$ and $||\mathbf{p}_{k}||^{2}=1$, $\forall k \in \mathcal{K}$. 
The transmitted streams satisfy \mbox{$\mathbb{E}\left\lbrace \mathbf{s}\mathbf{s}^{H}\right\rbrace =\mathbf{I}$}, where  \mbox{$\mathbf{s}=\left[s_{c}, s_{1}, \ldots, s_{k}\right]$}. The power allocation coefficient $t$ determines the distribution of power among the common precoder and the private precoders. 
Note that $t=1$ turns off the common stream and the strategy in \eqref{eqn:1} simply boils down to conventional SDMA with uniform power allocation.

The signal received by user-$k$ is written as
\begin{align}
	y_{k}=\mathbf{h}_{k}^{H}[m]\mathbf{x}+n_{k}, \quad k\in\mathcal{K},  
\end{align}
where $\mathbf{h}_{k}[m] \in \mathbb{C}^{n_{t}}$ is the channel vector between the transmitter and user-$k$ at time instant $m$ and $n_{k} \sim \mathcal{CN}(0,1)$ is the Additive White Gaussian Noise (AWGN) component for user-$k$. 

The flexible treatment\footnote{Readers are referred to \cite{clerckx_2016,mao_2018} for more discussion on rate-splitting.} of interference is enabled by the joint use of message split at the transmitter and SIC at the receivers.  
The SIC processing at an arbitrary receiver ({\sl e.g.,} user-$k$, $k=\left\lbrace1,2,\ldots, K \right\rbrace $) can be explained in the following $3$ steps. 
\begin{enumerate}
	\item First, the interference from all users is partially decoded by decoding the common message estimate $\widehat{W}_{c}$ while treating the private streams of all ($K$) users as noise. Since the common message contains the common part of the message of each user (including the common part of the message of user-$k$), each user partially decodes the message of another user (and therefore partially decodes interference).
	\item Then, the common stream is reconstructed using the common message estimate  $\widehat{W}_{c}$ and subtracted from the received signal by SIC to remove the decoded interference from the received signal.
	\item Finally, the remaining (private) part of the message of user-$k$ (private part) is decoded by treating the interference from the private streams of all other $K-1$ users as noise. 
\end{enumerate}

The overall decoded message of user-$k$, $\widehat{W}_{k}$, is reconstructed by combining the decoded common part, $\widehat{W}_{c,k}$, and the decoded private part, $\widehat{W}_{p,k}$.

The Signal-to-Interference-plus-Noise Ratio (SINR) for the common stream at user-$k$ is
\begin{align}
	\gamma_{c,k}=\frac{P(1-t)|\mathbf{h}_{k}^{H}[m]\mathbf{p}_{c}|^{2}}{1+\frac{Pt}{K}\sum_{j \in \mathcal{K}}|\mathbf{h}_{k}^{H}[m]\mathbf{p}_{i}|^{2}}.
\end{align}

The SINR for the private stream at user-$k$ is given by
\begin{align}
	\gamma_{k}=\frac{\frac{Pt}{K}|\mathbf{h}_{k}^{H}[m]\mathbf{p}_{k}|^{2}}{1+\frac{Pt}{K}\sum_{j \in \mathcal{K}, j \neq k}|\mathbf{h}_{k}^{H}[m]\mathbf{p}_{i}|^{2}}.
\end{align}

The ergodic rates of the common and private streams are expressed as 
\begin{align}
	R_{c}(t)&\hspace{-0.1cm}=\hspace{-0.1cm}\mathbb{E}\left\lbrace \hspace{-0.1cm}\log_{2}\hspace{-0.1cm}\left(\hspace{-0.1cm} 1\hspace{-0.1cm}+\hspace{-0.1cm}\min_{k\in\mathcal{K}}\left[\frac{P(1-t)|\mathbf{h}^{H}_{k}[m]\mathbf{p}_{c}|^{2}}{\hspace{-0.1cm}1\hspace{-0.1cm}+\hspace{-0.1cm}\frac{Pt}{K}\sum_{j\in\mathcal{K}}|\mathbf{h}^{H}_{k}[m]\mathbf{p}_{j}|^{2}} \right] \right) \hspace{-0.1cm}\right\rbrace, \nonumber \\ 
	R_{k}(t)&\hspace{-0.1cm}=\hspace{-0.1cm}\mathbb{E}\left\lbrace \hspace{-0.1cm}\log_{2}\hspace{-0.1cm}\left(\hspace{-0.1cm} 1\hspace{-0.1cm}+\hspace{-0.1cm} \frac{\frac{Pt}{K} |\mathbf{h}^{H}_{k}[m]\mathbf{p}_{k}|^{2}}{1+ \frac{Pt}{K} \sum_{j\in\mathcal{K}, j\neq k}|\mathbf{h}^{H}_{k}[m]\mathbf{p}_{j}|^{2}} \right) \hspace{-0.1cm}\right\rbrace,
	\label{eqn:rates}
\end{align}
where the expectations are defined over the user channels $\mathbf{h}_{k}$.
The ergodic sum-rate for the power allocation coefficient $t$ is given by
\begin{align}
	R_{\mathrm{RSMA}}(t)=R_{c}(t)+\sum_{k=1}^{K}R_{k}(t).
	\label{eqn:sumrate}
\end{align}

We consider the scenarios in which imperfect CSIT occurs due to user mobility and delay in CSI reports from the users. We adopt the channel model in \cite{kim_2011}, where the channel coefficients at time instant $m$ are expressed as 
\begin{align}
	\mathbf{h}_{k}[m]=\sqrt{\epsilon^{2}} \mathbf{h}_{k}[m\minus1] + \sqrt{1-\epsilon^{2}} \mathbf{e}_{k}[m].
	\label{eqn:channel}
\end{align}
In the expression, $\mathbf{h}_{k}[m]$ denotes the spatially uncorrelated Rayleigh flat fading channel with i.i.d. entries distributed according to $\mathcal{CN}(0,1)$\footnote{The assumption of identical second order statistics for user channels is for the brevity of the derivations and to obtain a closed-form solution for the optimal power allocation coefficient $t$. The derivations for the lower bound on the ergodic sum-rate can be extended to user channels with different second-order statistics following the same steps in Section~\ref{sec:lowerboundergodic}.}, $\epsilon=J_{0}(2\pi f_{D}T)$ denotes the time correlation coefficient obeying the Jakes' model \cite{proakis_2000} and $\mathbf{e}_{k}[m]$ has i.i.d. entries distributed according to $\mathcal{CN}(0,1)$. The parameter $f_{D}=vf_{c}/c$ is the maximum Doppler frequency for given carrier frequency $f_{c}$ and user speed $v$ ($c$ is the speed of light) and $T$ is the channel instantiation interval. 
In words, $\mathbf{h}_{k}[m]$ is the instantaneous channel vector observed at time instant $m$. However, the transmitter only has knowledge of the channel vector observed at time instant $m\minus1$, $\mathbf{h}_{k}[m\minus 1]$, due to the latency in CSI reports from the users ($T$ models the CSI report delay in seconds).
Consequently, the transmitter uses the CSIT $\mathbf{h}_{k}[m\minus 1]$ at time instant $m$ to calculate the precoders. We consider ZF precoders with equal power allocation for the private streams\footnote{Aside from being extremely common in literature (for example, see \cite{hao_2015,dai_2016,dai_2017,jaramilloramirez_2015} and the references therein), equal power allocation among precoders in MU-MIMO is employed in practical systems like 4G and 5G (for example, see \cite{lim_2013, liu_2017}).}, so that \mbox{$|\mathbf{h}_{j}^{H}[m\minus1]\mathbf{p}_{k}|=0$}, $\forall j \in \mathcal{K}\backslash k$. The common precoder is assumed to be a random beamformer independent of $\mathbf{h}_{k}[m\minus 1]$, $\mathbf{e}_{k}[m]$, and $\mathbf{p}_{k}$, $\forall k\in\mathcal{K}$, in order to be able to obtain tractable expressions for the rate of the common stream\footnote{In Section~\ref{sec:simulation}, we show that RSMA achieves significant sum-rate gain even with random beamformers for common streams.}. 
\vspace{-0.4cm}

\section{Lower Bound for the Ergodic Sum-Rate}
\label{sec:lowerboundergodic}
Our aim is to find a power allocation method that maximizes the ergodic sum-rate of RSMA in \eqref{eqn:sumrate} under imperfect CSIT due to mobility. As explained in Section~\ref{sec:system}, we consider the RSMA scheme with ZF beamformers for the private streams with a portion $Pt$ of the total power distributed equally among the users. The rest of the power, $P(1-t)$, is allocated to the common stream. 

We focus on deriving a lower bound for the sum-rate of 1-layer RSMA scheme in an effort to obtain a closed-form solution for the optimum power allocation coefficient $t_{opt}$.
We obtain the lower bound for an approximation of the ergodic sum-rate, which will be referred as the Approximated Ergodic Sum-Rate (AESR) for the rest of the paper, to obtain tractable expressions and a closed-form solution for the power allocation coefficient. 
As a result, the obtained lower bound is for the AESR 
\begin{align}
	\widetilde{R}_{\mathrm{RSMA}}(t)=\widetilde{R}_{c}(t)+\sum_{k=1}^{K}\widetilde{R}_{k}(t),	\label{eqn:approx_sumrate}
\end{align}
where $\widetilde{R}_{c}(t)$ and $\widetilde{R}_{k}(t)$ are Approximated Ergodic Rates (AERs), which are approximations for $R_{c}(t)$ and $R_{k}(t)$, respectively. The approximations are explained in detail in the following subsections. We demonstrate the tightness of the obtained lower bound by simulations in Section~\ref{sec:simulation}.  

We perform the derivation in two parts: the derivation of a lower bound for the sum of the AERs of the private streams, $\sum_{k=1}^{K}\widetilde{R}_{k}(t)$, and the derivation of a lower bound for the AER of the common stream, $\widetilde{R}_{c}(t)$. 

\subsection{Lower Bound for $\sum_{k=1}^{K}\widetilde{R}_{k}(t)$ }
\label{sec:lowerbound_private}
We assume that all users experience channels with the same second order statistics. Therefore, we first find a lower bound for the ergodic rate of user-$k$, $R_{k}(t)$, and generalize it to $K$ users to obtain the lower bound for $\sum_{k=1}^{K}R_{k}(t)$ \cite{hao_2015}.

We rewrite the expression for $R_{k}(t)$ in \eqref{eqn:rates} as  
\begin{align}
	R_{k}(t)&=\hspace{-0.1cm}\mathbb{E}\left\lbrace \log_{2}\hspace{-0.1cm}\left(\hspace{-0.1cm} 1\hspace{-0.1cm}+\hspace{-0.1cm} \frac{Pt}{K}\hspace{-0.1cm}\left( |\mathbf{h}^{H}_{k}[m]\mathbf{p}_{k}|^{2}\hspace{-0.1cm}+\hspace{-0.3cm}\sum_{j\in\mathcal{K}, j\neq k}\hspace{-0.3cm}|\mathbf{h}^{H}_{k}[m]\mathbf{p}_{j}|^{2} \hspace{-0.1cm}\right) \hspace{-0.1cm}\right) \hspace{-0.1cm}\right\rbrace  \nonumber\\
	& -\mathbb{E}\left\lbrace \log_{2}\left( 1+ \frac{Pt}{K}\sum_{j\in\mathcal{K}, j\neq k} |\mathbf{h}^{H}_{k}[m]\mathbf{p}_{j}|^{2} \right) \right\rbrace.  
	\label{eqn:rewrite_rkt}
\end{align}
Expression \eqref{eqn:rewrite_rkt} consists of the terms  $|\mathbf{h}_{k}^{H}[m]\mathbf{p}_{j}|^{2}$, for $k,j \in \mathcal{K}$. For tractability, we approximate the terms as 
\begin{align}
	\hspace{-0.25cm}|\mathbf{h}_{k}^{H}[m]\mathbf{p}_{k}|^{2} \hspace{-0.1cm}&\approx\hspace{-0.1cm} \epsilon^{2}|\mathbf{h}_{k}^{H}[m-1]\mathbf{p}_{k}|^{2}\hspace{-0.1cm}+\hspace{-0.1cm}(1-\epsilon^{2})|\mathbf{e}_{k}^{H}[m]\mathbf{p}_{k}|^{2}\hspace{-0.1cm},\label{eqn:dist1}\\
	\hspace{-0.25cm}|\mathbf{h}_{k}^{H}[m]\mathbf{p}_{j}|^{2}\hspace{-0.1cm}&\approx \epsilon^{2}|\mathbf{h}_{k}^{H}[m-1]\mathbf{p}_{j}|^{2}\hspace{-0.1cm}+\hspace{-0.1cm}(1-\epsilon^{2})|\mathbf{e}_{k}^{H}[m]\mathbf{p}_{j}|^{2} \nonumber \\
	&=(1-\epsilon^{2})|\mathbf{e}_{k}^{H}[m]\mathbf{p}_{j}|^{2}.
	\label{eqn:dist2}
\end{align}
The approximations in \eqref{eqn:dist1} and \eqref{eqn:dist2} are due to the neglected terms containing both $\mathbf{h}_{k}^{H}[m-1]$ and $\mathbf{e}_{k}[m]$ \cite{jaramilloramirez_2015,zhang_2009, zhang_2011}. 
Observing the expressions \eqref{eqn:dist1} and \eqref{eqn:dist2}, one needs the distributions of the 
random variables (r.v.s) \mbox{$|\mathbf{h}_{k}^{H}[m-1]\mathbf{p}_{k}|^{2}$} and \mbox{$|\mathbf{e}^{H}_{j}[m]\mathbf{p}_{k}|^{2}$} for \mbox{$j,k \in \mathcal{K}$}. 

With ZF beamforming, the precoder for a specific user is chosen to be orthogonal to the $K-1$ channels of the other users. Due to the isotropic nature of independent and identically distributed (i.i.d) Rayleigh fading, the
orthogonality reduces the degrees of freedom at the transmitter to $n_{t}-K+1$. Therefore, for ZF precoders with equal power allocation,
the term \mbox{$|\sqrt{2}\mathbf{h}_{k}^{H}[m-1]\mathbf{p}_{k}|^{2}$} has a chi-squared distribution with a degree of freedom $2(n_{t}-K+1)$, i.e.,
\mbox{$|\sqrt{2}\mathbf{h}_{k}^{H}[m-1]\mathbf{p}_{k}|^{2}\sim \chi^{2}_{2(n_{t}-K+1)}$}. Equivalently, \mbox{$|\mathbf{h}_{k}^{H}[m-1]\mathbf{p}_{k}|^{2}\sim \mathrm{Gamma}(n_{t}-K+1,1) $} \cite{jindal_2009,jindal_2011,jaramilloramirez_2015}.

The ZF precoder $\mathbf{p}_{k}$ is isotropically distributed and independent of the Gaussian error \mbox{$\mathbf{e}^{H}_{j}[m]$, $\forall j,k \in \mathcal{K}$}, leading to an exponentially distributed r.v. with unit mean, or equivalently   
\mbox{$|\mathbf{e}^{H}_{j}[m]\mathbf{p}_{k}| ^{2}\sim \mathrm{Gamma}(1,1)$}.
Assuming each interference term is independent, the term $\sum_{j\in\mathcal{K}, j\neq k}|\mathbf{e}^{H}_{k}[m]\mathbf{p}_{j}| ^{2}$ can be approximated by a random variable (r.v.) with the distribution \mbox{$\mathrm{Gamma}(K-1,1)$} \cite{jaramilloramirez_2015, zhang_2009}. 

We use the r.v.s and the distributions discussed above to obtain a lower bound for $\sum_{k=1}^{K}\widetilde{R}_{k}(t)$.  
We start the derivations from the first term in the expression \eqref{eqn:rewrite_rkt}. Lemma 1 gives an approximate distribution for the r.v. $\sum_{j\in\mathcal{K}}|\mathbf{h}^{H}_{k}[m]\mathbf{p}_{j}|^{2}$. 

\textit{Lemma 1:} 
The random variable $X=\sum_{j\in\mathcal{K}}|\mathbf{h}^{H}_{k}[m]\mathbf{p}_{j}|^{2}$ can be approximated by a random variable $\widetilde{X}$ with the distribution $\mathrm{Gamma}(\widehat{D},\widehat{\theta})$, where
\begin{align}
	\widehat{D}&=\frac{\left[ \epsilon^{2}(n_{t}+1)+(1-2\epsilon^{2})K\right] ^{2}}{\epsilon^{4}(n_{t}+1)+(1-2\epsilon^{2})K}, \nonumber \\
	\widehat{\theta}&=\frac{\epsilon^{4}(n_{t}+1)+(1-2\epsilon^{2})K}{\epsilon^{2}(n_{t}+1)+(1-2\epsilon^{2})K}.
	\label{eqn:dandtheta}
\end{align}

\textit{Proof:} See Appendix~\ref{appendix:lemma1}. \hspace{4.7cm}$\blacksquare$

We write an approximation $\widetilde{R}_{k}(t)$ for $R_{k}(t)$ using Lemma 1 and \eqref{eqn:dist2} with \eqref{eqn:rewrite_rkt} as
\begin{align}
	\hspace{-0.2cm}R_{k}(t)&\approx\widetilde{R}_{k}(t) \nonumber \\
	&=\mathbb{E}\left\lbrace \log_{2}\hspace{-0.1cm}\left(\hspace{-0.1cm} 1\hspace{-0.1cm}+\hspace{-0.1cm} \frac{Pt}{K}\widetilde{X} \right) \right\rbrace-\mathbb{E}\left\lbrace \log_{2}\left( 1+ \frac{Pt}{K}\widetilde{Z} \right) \right\rbrace,
	\label{eqn:approx_rk}
\end{align}
with \mbox{$\widetilde{Z}=(1-\epsilon^{2})\hspace{-0.1cm}\sum_{j\in\mathcal{K}, j\neq k}\hspace{-0.1cm}|\mathbf{e}_{k}^{H}[m]\mathbf{p}_{j}|^{2}$}. Proposition 1 gives a lower bound for $\widetilde{R}_{k}(t)$.

\textit{Proposition 1:} The AER $\widetilde{R}_{k}(t)$ is lower bounded by
\begin{align}
	\widetilde{R}_{k}(t) &\geq \log_{2}\hspace{-0.1cm}\left(1+\frac{P}{K}e^{\mu}t\right) \hspace{-0.05cm}-\hspace{-0.05cm}\log_{2}\hspace{-0.1cm}\left(\hspace{-0.1cm}1\hspace{-0.1cm}+\hspace{-0.1cm}\frac{(K-1)(1-\epsilon^{2})P}{K}t\right)\hspace{-0.1cm},
\end{align}
where $\mu=\ln(\widehat{\theta}) +\Gamma^{\prime}(\widehat{D})/\Gamma(\widehat{D})$ and the variables $\widehat{D}$ and $\widehat{\theta}$ are given in \eqref{eqn:dandtheta}.

\textit{Proof:} 
We start by writing a lower bound for $\widetilde{R}_{k}(t)$ as
\begin{subequations}
	\begin{align} 
		\widetilde{R}_{k}(t) & \geq \mathbb{E}\hspace{-0.05cm}\left\lbrace\hspace{-0.05cm} \log_{2}\hspace{-0.05cm}\left(\hspace{-0.1cm}1\hspace{-0.05cm}+\hspace{-0.05cm} \frac{Pt}{K}\widetilde{X} \hspace{-0.05cm}\right)\hspace{-0.05cm} \right\rbrace \hspace{-0.05cm}-\hspace{-0.05cm}\log_{2}\hspace{-0.1cm}\left(\hspace{-0.05cm} 1\hspace{-0.05cm}+\hspace{-0.05cm}\frac{Pt}{K}\mathbb{E}\{\widetilde{Z}\}\right) 
		\label{eqn:jensens1} \\
		& \geq \log_{2}\hspace{-0.1cm}\left(\hspace{-0.1cm}1\hspace{-0.05cm}+\hspace{-0.05cm}\frac{Pt}{K}e^{\mathbb{E}\{\ln(\widetilde{X})\} } \hspace{-0.1cm}\right)\hspace{-0.05cm}-\hspace{-0.05cm}\log_{2}\hspace{-0.1cm}\left(\hspace{-0.1cm}1\hspace{-0.05cm}+\hspace{-0.05cm}\frac{Pt}{K}\mathbb{E}\{\widetilde{Z}\}\hspace{-0.1cm}\right). \label{eqn:jensens2}
	\end{align}
	\label{eqn:jensens}
\end{subequations}
\hspace{-0.1cm}The expressions \eqref{eqn:jensens1} and \eqref{eqn:jensens2} follow from the Jensen's inequality and the fact that $\log_{2}(1+ax)$ is a concave function of $x$ for any $ax>0$ and $\log_{2}(1+ae^{x})$ is a convex function of $x$ for any $a>0$ \cite{hao_2015}, respectively. 

The mean of $\ln(\widetilde{X})$ can be shown to be
\begin{align}
	\mu\triangleq\mathbb{E}\{\ln(\widetilde{X})\}=\ln(\widehat{\theta})+\Gamma^{\prime}(\widehat{D})/\Gamma(\widehat{D}).
\end{align} 
As a result,
\begin{align}
	\log_{2}\hspace{-0.1cm}\left(1+\frac{Pt}{K}e^{\mathbb{E}\left\lbrace\ln \widetilde{X} \right\rbrace} \right)=\log_{2}\left(1+\frac{Pt}{K}e^{\mu}\right). \nonumber
\end{align}
For the second term in \eqref{eqn:approx_rk}, we can directly write \mbox{$\mathbb{E}\{\widetilde{Z}\}=(1-\epsilon^{2})(K-1)$} as
$$\sum_{j\in\mathcal{K}, j\neq k}|\mathbf{e}^{H}_{k}[m]\mathbf{p}_{j}|^{2} \sim \mathrm{Gamma}(K-1,1).$$ 
Therefore, 
\begin{align}
	\log_{2}\left(1+\frac{Pt}{K}\mathbb{E}\{\widetilde{Z}\}\right)=\log_{2}\left( 1+\frac{P(1-\epsilon^{2})(K-1)}{K}t\right), \nonumber
\end{align}
which completes the proof. \hspace{4.5cm}$\blacksquare$

We can generalize the lower bound given in Proposition 1 to $K$ users with independent channels having identical second order statistics as 
\begin{align}
	\sum_{k=1}^{K}\widetilde{R}_{k}(t) &\geq K\log_{2}\left(1+\frac{P}{K}e^{\mu}t\right) \nonumber \\ &\quad \quad\quad   -K\log_{2}\left(\hspace{-0.1cm}1\hspace{-0.1cm}+\hspace{-0.1cm}\frac{(K-1)(1-\epsilon^{2})P}{K}t\right).
\end{align} 

\subsection{Lower Bound for $\widetilde{R}_{c}(t)$}
\label{sec:lowerbound}
Observing the expression for $R_{c}(t)$ in \eqref{eqn:rates}, we define the r.v.s
\begin{align}
	Y_{k}&=\frac{|\mathbf{h}^{H}_{k}[m]\mathbf{p}_{c}|^{2}}{1+\frac{Pt}{K}\sum_{j\in\mathcal{K}} |\mathbf{h}^{H}_{k}[m]\mathbf{p}_{j}|^{2}},
\end{align}
and
\begin{align}
	Y=\min_{k\in\mathcal{K}}Y_{k}.
\end{align}

For tractability of the expressions, we approximate the r.v. $Y$ by a r.v. $\widetilde{Y}$, so that
\begin{align}
	\widetilde{R}_{c}(t)=\log_{2}\left(1+P(1-t)\widetilde{Y}\right).
\end{align}

The following lemma gives the distribution of $\widetilde{Y}$.

\textit{Lemma 2:} The r.v. $Y$ can be approximated by a r.v. $\widetilde{Y}$ with the CDF 
\begin{align}
	F_{\widetilde{Y}}(y)&\approx 1-\frac{e^{-Ky}}{\left(y\frac{P\widehat{\theta}t}{K}+1 \right)^{\widehat{D}K} }.
	\label{eqn:cdf_Y}
\end{align}

\textit{Proof:} See Appendix~\ref{appendix:lemma2}.\hspace{4.7cm}$\blacksquare$

Before we start to derive the lower bound, we give another useful result in Lemma 3. 

\textit{Lemma 3:} The expected value of the natural logarithm of the r.v. $\widetilde{Y}$ is given by
\begin{align}
	\mathbb{E}\{\ln(\widetilde{Y})\}&=-\gamma-\ln(K)-e^{\frac{K^{2}}{P\widehat{\theta}t}}\sum_{m=1}^{\left \lfloor{\widehat{D}K}\right \rceil}\mathrm{E}_{m}\left( \frac{K^{2}}{P\widehat{\theta}t}\right) ,\end{align}
where $\gamma\approx0.577$ is the Euler-Mascheroni constant and
\begin{align}
	E_{v}(x)=\int_{1}^{\infty}\frac{e^{-tx}}{t^{v}}dt, \quad x > 0, \ v \in \mathbb{R}
\end{align}
is the generalized exponential-integral function \cite{milgram_1985,chiccoli_1990}.

\textit{Proof:} The expected value of the function $g(x)$ of the random variable $x$ can be calculated from the CDF $F_{X}(x)$ as
\begin{align}
	&\mathbb{E}\left\lbrace g(x)\right\rbrace=\int_{-\infty}^{\infty} g(x)\frac{dF_{X}(x)}{dx}=-\int_{-\infty}^{\infty} g(x)\frac{d(1-F_{X}(x))}{dx}.
\end{align}
Integrating by parts gives
\begin{align}
	\mathbb{E}\left\lbrace g(x)\right\rbrace\hspace{-0.1cm}=\hspace{-0.1cm}- g(x)(1\hspace{-0.1cm}-\hspace{-0.1cm}F_{X}(x))\bigg\rvert_{-\infty}^{\infty}\hspace{-0.35cm} +\int_{-\infty}^{\infty}\hspace{-0.35cm}(1\hspace{-0.1cm}-\hspace{-0.1cm}F_{X}(x))\frac{d(g(x))}{dx}.
	\label{eqn:expectation}
\end{align}

We use the property in \eqref{eqn:expectation} and the CDF for $\ln(\widetilde{Y})$ given in \eqref{eqn:cdf_Y} to obtain $\mathbb{E}\{\ln(\widetilde{Y})\}$ as
\begin{align}
	&\mathbb{E}\{\ln(\widetilde{Y})\}\nonumber \\
	&=-\ln(y)\frac{e^{-Ky}}{\left(y\frac{P\widehat{\theta}t}{K}+1 \right)^{\widehat{D}K} }\bigg\rvert_{0}^{\infty}+\int_{0}^{\infty}\frac{e^{-Ky}}{y\left(y\frac{P\widehat{\theta}t}{K}+1 \right)^{\widehat{D}K} }dy. \nonumber \\
\end{align}

We assume that the power allocated to the private streams is always non-zero, so that $\frac{P\widehat{\theta}t}{K}>0$. Under the assumption, we perform a change of variables in the second part of the expression by $z=\left(y\frac{P\widehat{\theta}t}{K}+1 \right) $, so that,
\begin{align}
	&\mathbb{E}\{\ln(\widetilde{Y})\} \nonumber \\
	&=-\ln(y)\frac{e^{-Ky}}{\left(y\frac{P\widehat{\theta}t}{K}+1 \right)^{\widehat{D}K} }\bigg\rvert_{0}^{\infty}+\int_{1}^{\infty}\frac{e^{-K(z-1)\frac{K}{P\widehat{\theta}t}}}{z^{\widehat{D}K}\frac{z-1}{\left( \frac{P\widehat{\theta}t}{K}\right) }}\frac{dz}{\left( \frac{P\widehat{\theta}t}{K}\right) } \nonumber \\
	&=-\ln(y)\frac{e^{-Ky}}{\left(y\frac{P\widehat{\theta}t}{K}+1 \right)^{\widehat{D}K} }\bigg\rvert_{0}^{\infty} +e^{\frac{K^{2}}{P\widehat{\theta}t}}\int_{1}^{\infty}\frac{e^{-\frac{K^{2}}{P\widehat{\theta}t}z}}{z^{\widehat{D}K}(z-1)}dz \nonumber \\
	&=-\ln(y)\frac{e^{-Ky}}{\left(y\frac{P\widehat{\theta}t}{K}+1 \right)^{\widehat{D}K} }\bigg\rvert_{0}^{\infty}+e^{\frac{K^{2}}{P\widehat{\theta}t}}\Theta(\widehat{D}K).
	\label{eqn:nontractable1}
\end{align}
The integral $\Theta(\widehat{D}K)$ is not tractable. Our aim is to transform it into a form that is expressed in terms of known integrals. We use the expansion $\frac{1}{z(z-1)}=\frac{1}{z-1}-\frac{1}{z}$ to write
\begin{align}
	\Theta(\widehat{D}K)&=\int_{1}^{\infty}\frac{e^{-\frac{K^{2}}{P\widehat{\theta}t} z}}{z^{\widehat{D}K}(z-1)}dz \nonumber \\
	&=\int_{1}^{\infty}\frac{e^{-\frac{K^{2}}{P\widehat{\theta}t} z}}{z^{(\widehat{D}K-1)}z(z-1)}dz\nonumber \\
	&=\int_{1}^{\infty}\frac{e^{-\frac{K^{2}}{P\widehat{\theta}t} z}}{z^{(\widehat{D}K-1)}}\left(\frac{1}{z-1}-\frac{1}{z} \right)dz \nonumber \\
	&=\int_{1}^{\infty}\frac{e^{-\frac{K^{2}}{P\widehat{\theta}t} z}}{z^{(\widehat{D}K-1)}(z-1)}dz-\int_{1}^{\infty}\frac{e^{-\frac{K^{2}}{P\widehat{\theta}t} z}}{z^{\widehat{D}K}}dz \nonumber \\
	&=\Theta(\widehat{D}K-1)-\mathrm{E}_{\widehat{D}K}\left( \frac{K^{2}}{P\widehat{\theta}t}\right).
	\label{eqn:recursion}
\end{align}
In order to define a tractable recursive expression to solve \eqref{eqn:recursion}, we need $\widehat{D}K=\frac{\left[ \epsilon^{2}(n_{t}+1)+(1-2\epsilon^{2})K\right] ^{2}}{\epsilon^{4}(n_{t}+1)/K+(1-2\epsilon^{2})}$ to be a positive integer. Since the multiplication is not guaranteed to give such a result,
we replace $\widehat{D}K$ with $\left \lfloor{\widehat{D}K}\right \rceil $. We rewrite the expression in \eqref{eqn:recursion} as
\begin{align}
	\Theta\left( \left \lfloor{\widehat{D}K}\right \rceil\right) &=\Theta\left( \left \lfloor{\widehat{D}K}\right \rceil-1\right) -\mathrm{E}_{\left \lfloor{\widehat{D}K}\right \rceil}\left( \frac{K^{2}}{P\widehat{\theta}t}\right) \nonumber \\
	&=\Theta(0) -\sum_{m=1}^{\left \lfloor{\widehat{D}K}\right \rceil}\hspace{-0.1cm}\mathrm{E}_{m}\hspace{-0.1cm}\left(\hspace{-0.04cm} \frac{K^{2}}{P\widehat{\theta}t}\hspace{-0.04cm}\right).
	\label{eqn:recursion_remastered}
\end{align}

One needs a closed-form solution for $\Theta(0)=\int_{1}^{\infty}\frac{e^{-\frac{K^{2}}{P\widehat{\theta}t} z}}{z-1}dz$ in order to obtain a closed-form expression for \eqref{eqn:recursion_remastered}. For this purpose, we perform a change of variables $u=\frac{K^{2}}{P\widehat{\theta}t} z-\frac{K^{2}}{P\widehat{\theta}t}$, so that
\begin{align}
	\Theta(0)=e^{-\frac{K^{2}}{P\widehat{\theta}t}}\int_{0}^{\infty}\frac{e^{-u}}{u}du=e^{-\frac{K^{2}}{P\widehat{\theta}t}}\int_{0}^{\infty}\hspace{-0.2cm}e^{-u}\frac{d(\ln (u))}{du}.
\end{align}
Next, we perform integration by parts to obtain
\begin{subequations}
	\begin{align}
		\Theta(0)&=e^{-\frac{K^{2}}{P\widehat{\theta}t}}e^{-u}\ln(u)\bigg\rvert_{0}^{\infty}\hspace{-0.2cm}+e^{-\frac{K^{2}}{P\widehat{\theta}t}}\int_{0}^{\infty}e^{-u}\ln(u)du  \\
		&=e^{-\frac{K^{2}}{P\widehat{\theta}t}}e^{-u}\ln(u)\bigg\rvert_{0}^{\infty}-e^{-\frac{K^{2}}{P\widehat{\theta}t}}\gamma  \\
		&=e^{-\frac{K^{2}}{P\widehat{\theta}t}}\left( e^{-Ky}\ln(y)\bigg\rvert_{0}^{\infty}-\ln(K)-\gamma\right). \label{eqn:lnk} 
	\end{align}
\end{subequations}
The expression \eqref{eqn:lnk} follows from the fact that $u=Ky$. We solve the recursive expression in \eqref{eqn:recursion_remastered} by substituting \eqref{eqn:lnk} to get
\begin{align}
	&\Theta\left( \left \lfloor{\widehat{D}K}\right \rceil\right) \nonumber \\ 
	&=e^{-\frac{K^{2}}{P\widehat{\theta}t}}\hspace{-0.1cm}\left(\hspace{-0.1cm} e^{-Ky}\ln(y)\bigg\rvert_{0}^{\infty}\hspace{-0.3cm}-\ln(K)-\gamma\hspace{-0.05cm}\right) \hspace{-0.1cm}-\hspace{-0.2cm}\sum_{m=1}^{\left \lfloor{\widehat{D}K}\right \rceil}\hspace{-0.2cm}\mathrm{E}_{m}\hspace{-0.1cm}\left( \frac{K^{2}}{P\widehat{\theta}t}\right).
	\label{eqn:recursion_solved} 
\end{align}

Finally, we substitute the solution in \eqref{eqn:recursion_solved} into \eqref{eqn:nontractable1} to get
\begin{align}
	&\mathbb{E}\{\ln(\widetilde{Y})\} \nonumber \\
	&=-\ln(y)\frac{e^{-Ky}}{\left(y\frac{P\widehat{\theta}t}{K}+1 \right)^{\widehat{D}K} }\bigg\rvert_{0}^{\infty}\hspace{-0.25cm}+\ln(y)e^{-Ky}\bigg\rvert_{0}^{\infty}\hspace{-0.25cm}-\hspace{-0.05cm}\ln(\hspace{-0.03cm}K\hspace{-0.03cm})\hspace{-0.05cm}-\hspace{-0.05cm}\gamma\hspace{-0.05cm}\nonumber\\
	&\hspace{5.5cm}-e^{\frac{K^{2}}{P\widehat{\theta}t}}\hspace{-0.1cm}\sum_{m=1}^{\left \lfloor{\widehat{D}K}\right \rceil}\hspace{-0.1cm}\mathrm{E}_{m}\hspace{-0.1cm}\left(\hspace{-0.04cm} \frac{K^{2}}{P\widehat{\theta}t}\hspace{-0.04cm}\right)   \nonumber\\
	&=\ln(y)e^{-Ky}\left( 1-\frac{1}{\left(y\frac{P\widehat{\theta}t}{K}+1 \right)^{\widehat{D}K} }\right) \bigg\rvert_{0}^{\infty}\hspace{-0.25cm}-\hspace{-0.05cm}\ln(\hspace{-0.03cm}K\hspace{-0.03cm})\hspace{-0.05cm}-\hspace{-0.05cm}\gamma\hspace{-0.05cm}\nonumber\\
	&\hspace{5.5cm}-e^{\frac{K^{2}}{P\widehat{\theta}t}}\hspace{-0.1cm}\sum_{m=1}^{\left \lfloor{\widehat{D}K}\right \rceil}\hspace{-0.1cm}\mathrm{E}_{m}\hspace{-0.1cm}\left(\hspace{-0.04cm} \frac{K^{2}}{P\widehat{\theta}t}\hspace{-0.04cm}\right)   \nonumber\\
	&=-\gamma-\ln(\hspace{-0.03cm}K\hspace{-0.03cm})-e^{\frac{K^{2}}{P\widehat{\theta}t}}\hspace{-0.1cm}\sum_{m=1}^{\left \lfloor{\widehat{D}K}\right \rceil}\hspace{-0.1cm}\mathrm{E}_{m}\hspace{-0.1cm}\left(\hspace{-0.04cm} \frac{K^{2}}{P\widehat{\theta}t}\hspace{-0.04cm}\right) . \nonumber \hspace{2.9cm}\blacksquare
\end{align}

\textit{Proposition 2:} The AER $\widetilde{R}_{c}(t)$ is lower bounded by
\begin{align}
	\widetilde{R}_{c}(t)\geq\log_{2}\left(1+P(1-t)e^{\beta} \right),
	\label{eqn:proposition2}
\end{align}
where $\beta\triangleq -\gamma-\ln(K)-e^{\frac{K^{2}}{P\widehat{\theta}t}}\sum_{m=1}^{\left \lfloor{\widehat{D}K}\right \rceil}\mathrm{E}_{m}\left(\frac{K^{2}}{P\widehat{\theta}t}\right) $.

\textit{Proof:} 
The proof is straightforward using the Jensen's inequality as in \eqref{eqn:jensens2} and the result in Lemma 3.  \hspace{1.3cm}$\blacksquare$

The results in Propositions~1~and~2 are combined to obtain a lower bound for an approximation of the sum-rate of the 1-layer RSMA with imperfect CSIT due to mobility.  

\textit{Proposition 3:} The ergodic sum-rate $R_{\mathrm{RSMA}}(t)$ of the 1-layer RSMA with imperfect CSIT due to mobility can be approximated by $\widetilde{R}_{\mathrm{RSMA}}(t)$, which is lower bounded by
\begin{align}
	&\widetilde{R}_{\mathrm{RSMA}}(t) \nonumber \\
	&\geq K\log_{2}\hspace{-0.1cm}\left(1+\frac{P}{K}e^{\mu}t\right) \hspace{-0.05cm}-\hspace{-0.05cm}K\log_{2}\hspace{-0.1cm}\left(\hspace{-0.1cm}1\hspace{-0.1cm}+\hspace{-0.1cm}\frac{(K-1)(1-\epsilon^{2})P}{K}t\right)\nonumber\\
	&\hspace{3.55cm}+\log_{2}\left(1+P(1-t)e^{\beta} \right). 
	\label{eqn:lowerbound}
\end{align}

We note here that the lower bound on the ergodic sum-rate can be extended to scenarios including user channels with different second-order statistics following the same steps in this section.

\section{Closed-Form Solution for the Optimal Power Allocation}
\label{sec:closedform}
We are interested in the value of the power allocation coefficient $t$, which maximizes the lower bound in \eqref{eqn:lowerbound}. Such a coefficient can be found by an exhaustive search over the expression \eqref{eqn:lowerbound}. However, obtaining the optimum $t$ by an exhaustive search is not suitable for use in practical systems. In this section, we aim to find a closed-form solution for the optimal power allocation coefficient, which we denote as $t_{opt}$. 

\subsection{Derivation}
We assume the asymptotic scenario where $Pt \rightarrow \infty$ in an effort to obtain the closed-form solution. With such assumption, we obtain the following result.

\textit{Lemma 4:} As $Pt \rightarrow \infty$, the term $\beta=-\gamma-\ln(K)-e^{\frac{K^{2}}{P\widehat{\theta}t}}\sum_{m=1}^{\left \lfloor{\widehat{D}K}\right \rceil}\mathrm{E}_{m}\left(\frac{K^{2}}{P\widehat{\theta}t}\right) $ can be approximated as
\begin{align}
	&\beta\approx-\gamma\hspace{-0.05cm}+\hspace{-0.05cm}\ln\hspace{-0.05cm}\left(\frac{K}{P\widehat{\theta}\left( \left \lfloor{\widehat{D}K}\right \rceil-1\right) t} \right)\hspace{-0.05cm}-\hspace{-0.05cm}\frac{1}{2(\floor{\widehat{D}K}-1)}.
	\label{eqn:beta_approx0}
\end{align} 

\textit{Proof:} See Appendix~\ref{appendix:lemma4}. \hspace{4.7cm}$\blacksquare$

Based on the assumption $Pt \rightarrow \infty$ and Lemma 4, we can rewrite the lower bound expression \eqref{eqn:lowerbound} as
\begin{align}
	&\widetilde{R}_{\mathrm{RSMA}}(t) \nonumber \\
	&\geq K\log_{2}\hspace{-0.1cm}\left(1+\frac{Pe^{\mu}}{K}t\right) \hspace{-0.05cm}-\hspace{-0.05cm}K\log_{2}\hspace{-0.1cm}\left(\hspace{-0.1cm}1\hspace{-0.1cm}+\hspace{-0.1cm}\frac{(K-1)(1-\epsilon^{2})P}{K}t\right)\nonumber \\&\hspace{5.0cm}+\log_{2}\left(1+P(1-t)e^{\beta} \right) \nonumber \\
	&\approx K\log_{2}\left(\frac{Pe^{\mu}}{K}t\right)-K\log_{2}\left(\hspace{-0.1cm}1\hspace{-0.1cm}+\hspace{-0.1cm}\frac{(K-1)(1-\epsilon^{2})P}{K}t\right)\nonumber \\
	&\hspace{1.5cm}+\log_{2}\left(\hspace{-0.1cm}1\hspace{-0.1cm}+\hspace{-0.1cm}\frac{(1-t)}{t}\frac{K}{\hat{\theta}(\left \lfloor{\widehat{D}K}\right \rceil-1)}e^{-\gamma-\frac{1}{2(\left \lfloor{\widehat{D}K}\right \rceil-1)}} \right)\nonumber \\
	& = -K\log_{2}\left(\frac{1}{\tau t}+\frac{\omega}{\tau}\right)+\log_{2}\left((1-\rho)+\frac{\rho}{t}\right).
	\label{eqn:simplified}
\end{align}
with the terms
\begin{align}
	\tau&\triangleq\frac{Pe^{\mu}}{K}, \quad\quad \omega\triangleq \frac{(K-1)(1-\epsilon^{2})P}{K}, \nonumber\\ 
	\rho& \triangleq\frac{K}{\widehat{\theta}(\left \lfloor{\widehat{D}K}\right \rceil-1)}e^{-\gamma-\frac{1}{2(\left \lfloor{\widehat{D}K}\right \rceil-1)}}.
	\label{eqn:terms}
\end{align} 

Taking the derivate of \eqref{eqn:simplified} with respect to $t$ and equating the resulting expression to zero yields
\begin{figure*}[t!]
	\begin{subfigure}{.5\textwidth}
		\centerline{\includegraphics[width=3.2in,height=3.2in,keepaspectratio]{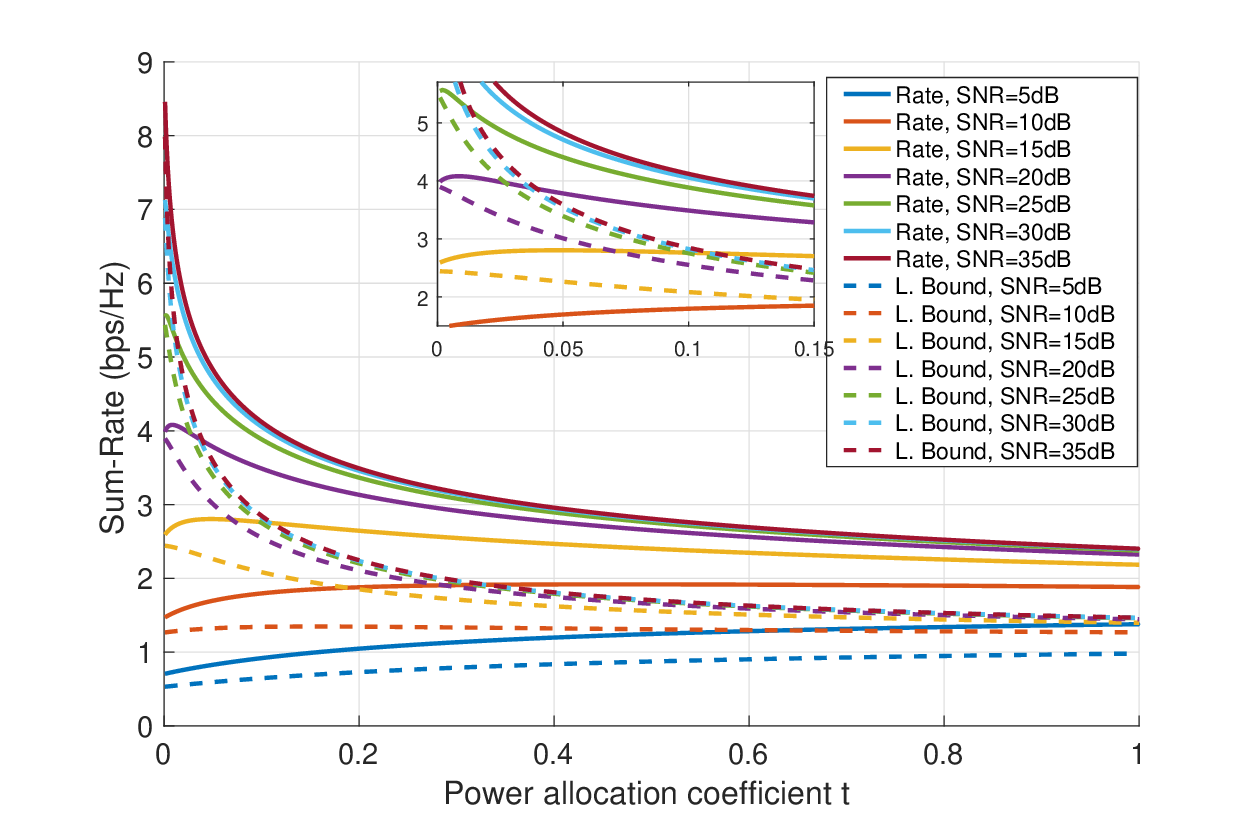}}
		\caption{$K=4$, $n_{t}=4$, $\epsilon=0.5$.}
		\label{fig:lowerbound1}
	\end{subfigure}
	\begin{subfigure}{.5\textwidth}
		\centerline{\includegraphics[width=3.2in,height=3.2in,keepaspectratio]{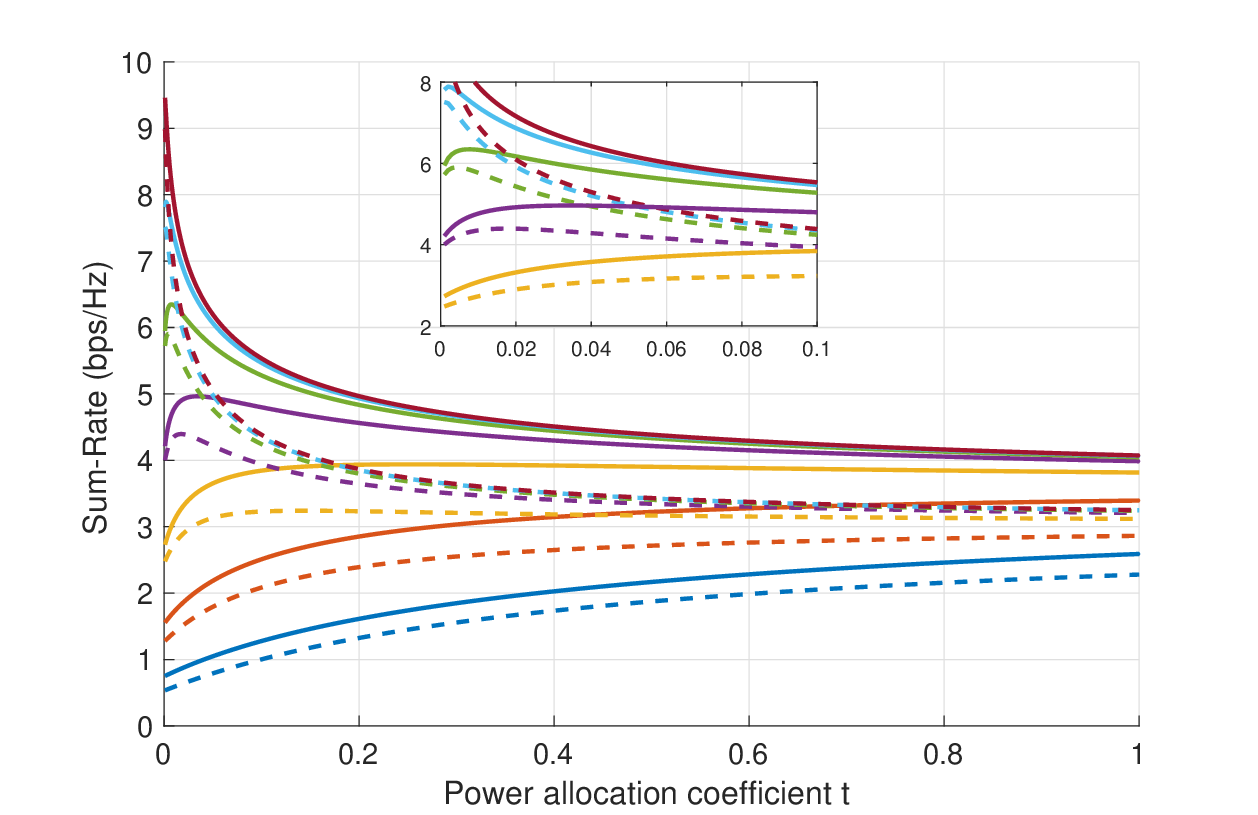}}
		\caption{$K=4$, $n_{t}=8$, $\epsilon=0.5$.}
		\label{fig:lowerbound2}
	\end{subfigure}
	\newline
	\begin{subfigure}{.5\textwidth}
		\centerline{\includegraphics[width=3.2in,height=3.2in,keepaspectratio]{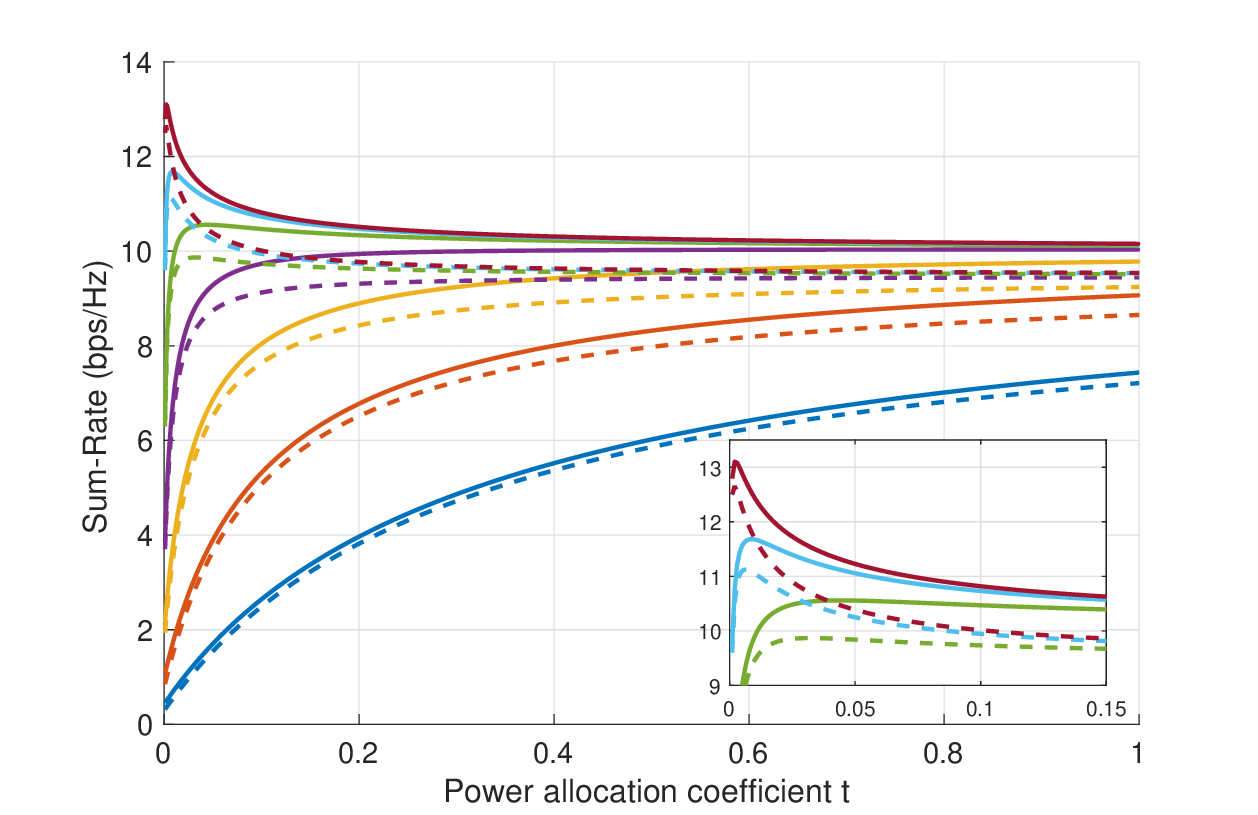}}
		\caption{$K=8$, $n_{t}=32$, $\epsilon=0.5$.}
		\label{fig:lowerbound7}
	\end{subfigure}
	\begin{subfigure}{.5\textwidth}
		\centerline{\includegraphics[width=3.2in,height=3.2in,keepaspectratio]{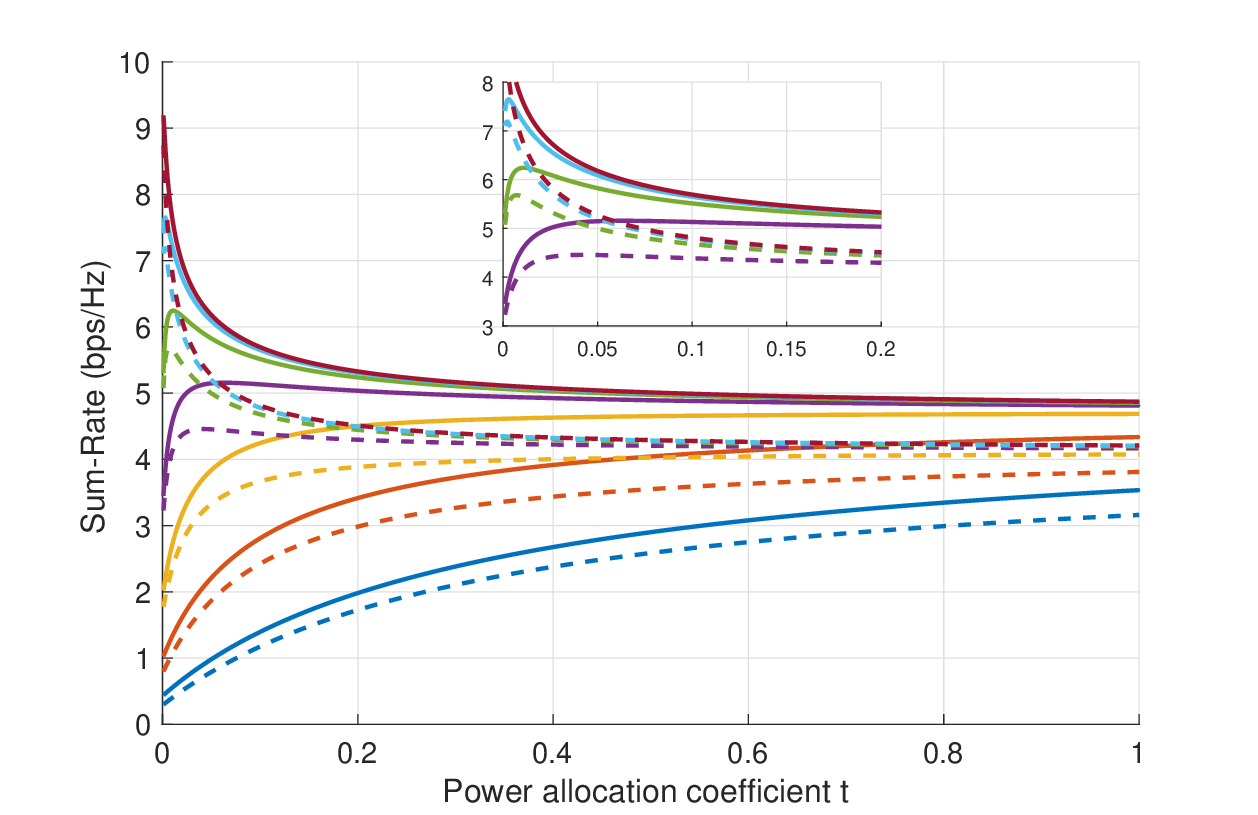}}
		\caption{$K=8$, $n_{t}=32$, $\epsilon=0.3$.}
		\label{fig:lowerbound8}
	\end{subfigure}
	\caption{Sum-rate and lower bound.}
	\vspace{-0.5cm}
\end{figure*}
\begin{align}
	t_{opt}=\frac{\rho(K-1)}{\rho(\omega+K)-K}.
\end{align}

In order to obtain a feasible value for $t_{opt}$, i.e., $t_{opt} \in \left( 0,1\right] $, we set the condition
$\rho(\omega+K)-K > \rho(K-1)$. The resulting power allocation algorithm is expressed as
\begin{align}
	t_{opt}=
	\left\{
	\begin{array}{ll}
		\frac{\rho(K-1)}{\rho(\omega+K)-K},  &  \mbox{if} \ \rho(\omega+1)/K > 1\\
		\hspace{0.8cm} 1\hspace{0.6cm}, & \mbox{otherwise}.
	\end{array}
	\right.
	\label{eq:poweralloc}
\end{align} 

In the next subsection, we analyze the expression \eqref{eq:poweralloc} for extreme cases of CSIT quality \mbox{(i.e., $\epsilon=0$ and $\epsilon=1$)} in order to get an insight on the behaviour of the proposed power allocation method.

\subsection{Asymptotic Analysis}
We analyze the values of the power allocation coefficient for the asymptotic values of $0$ and $1$ for $\epsilon\in \left[0,1 \right] $. 
As the first step in the analyses, we check whether the condition in \eqref{eq:poweralloc} is feasible (i.e., can be satisfied for at least one combination of $K$, $n_{t}$ and $P$) in the considered asymptotic case.
\subsubsection{$\epsilon \rightarrow $ 0} We substitute $\epsilon = 0$ in \eqref{eqn:dandtheta} and \eqref{eqn:terms} to obtain the following: $\hat{D} = K$, $\hat{\theta} = 1$, $\omega = \frac{P(K-1)}{K}$, and $\rho = \frac{K}{(K^{2}-1)}e^{-\gamma-\frac{1}{2(K^{2}-1)}}$.
Then, the condition in \eqref{eq:poweralloc} becomes 
\begin{align}
	1 < \frac{P(1-\frac{1}{K})+1}{K^{2}-1}e^{-\gamma-\frac{1}{2(K^{2}-1)}}.
	\label{eqn:cond1}
\end{align}
The expression \eqref{eqn:cond1} is difficult to analyze in its current form. We divert our analysis to the scenarios with the smallest possible number ($K=2$) and a large ($K-1\approx K$) number of users in an effort to obtain an understanding for the coefficient behaviour at extreme values.  

For $K=2$, the power allocation algorithm becomes  
\begin{align}
	t_{opt}=
	\left\{
	\begin{array}{ll}
		\frac{2}{P-8.62},  & \quad \mbox{if} \ 10.62 < P \\
		\hspace{0.4cm} 1\hspace{0.6cm},  &\quad \mbox{otherwise}.
	\end{array}
	\right.
	\label{eq:poweralloc2}
\end{align}
It is seen from \eqref{eq:poweralloc2} that the resulting power allocation algorithm is dependent only on the transmit power level. The optimal power allocation coefficient
$t_{opt} \rightarrow 0$ as $P\rightarrow \infty$.

For large values of $K$, i.e., $K-1\approx K$, the power allocation condition is approximated as
$$
1.78 < \frac{P+1}{K^{2}}e^{-\frac{1}{2K^{2}}}\approx\frac{P+1}{K^{2}}.
$$
Then, the power allocation algorithm becomes 
\begin{align}
	t_{opt}=
	\left\{
	\begin{array}{ll}
		\frac{1}{\frac{P}{K}+1-1.78K},  & \quad \mbox{if} \ 1.78 < \frac{P+1}{K^{2}} \\
		\hspace{0.4cm} 1\hspace{1.15cm},  &\quad \mbox{otherwise}.
	\end{array}
	\right.
	\label{eq:poweralloc3}
\end{align}
The algorithm in \eqref{eq:poweralloc3} is similar to that in \eqref{eq:poweralloc2} in the sense that both rely on the transmit power level and the convergence behaviour of the optimal coefficient, i.e., $t_{opt} \rightarrow 0$ as $P\rightarrow \infty$.

The resulting algorithms indicate that at asymptotically low CSIT quality, more transmit power is assigned to the common stream as the transmit power increases. The threshold for power allocation to the common stream depends on the transmit power and the number of users. Such behaviour is justified since assigning more power (and equally distributing it) to the private streams does not improve the sum-rate in the interference-limited regime.

\subsubsection{$\epsilon \rightarrow $1} We substitute $\epsilon = 1$ in \eqref{eqn:dandtheta} and \eqref{eqn:terms} to obtain the following: \mbox{$\hat{D} = n_{t}+1-K$}, \mbox{$\hat{\theta} = 1$}, \mbox{$\omega = 0$}, and \mbox{$\rho = \frac{K}{(n_{t}+1-K)K-1}e^{-\gamma-\frac{1}{2((n_{t}+1-K)K-1)}}$}. 
Then, the condition in \eqref{eq:poweralloc} becomes 
\begin{align}
	1 < \frac{1}{(n_{t}+1-K)K-1}e^{-\gamma-\frac{1}{2((n_{t}+1-K)K-1)}}.
	\label{eqn:cond2}
\end{align}
As $e^{-\gamma-\frac{1}{2((n_{t}+1-K)K-1)}} < 1$, $K>1$, and $n_{t} \geq K$, the condition is not satisfied for any values of $K$ or $n_{t}$. Therefore, the optimal power allocation coefficient becomes $t_{opt}=1$, which indicates that no power is allocated to the common stream when there is perfect CSIT. Such behaviour is justified since multi-user interference can be cancelled out perfectly by ZF precoding under perfect CSIT.
\begin{figure*}[t!]
	\begin{subfigure}{.5\textwidth}
		\centerline{\includegraphics[width=3.1in,height=3.1in,keepaspectratio]{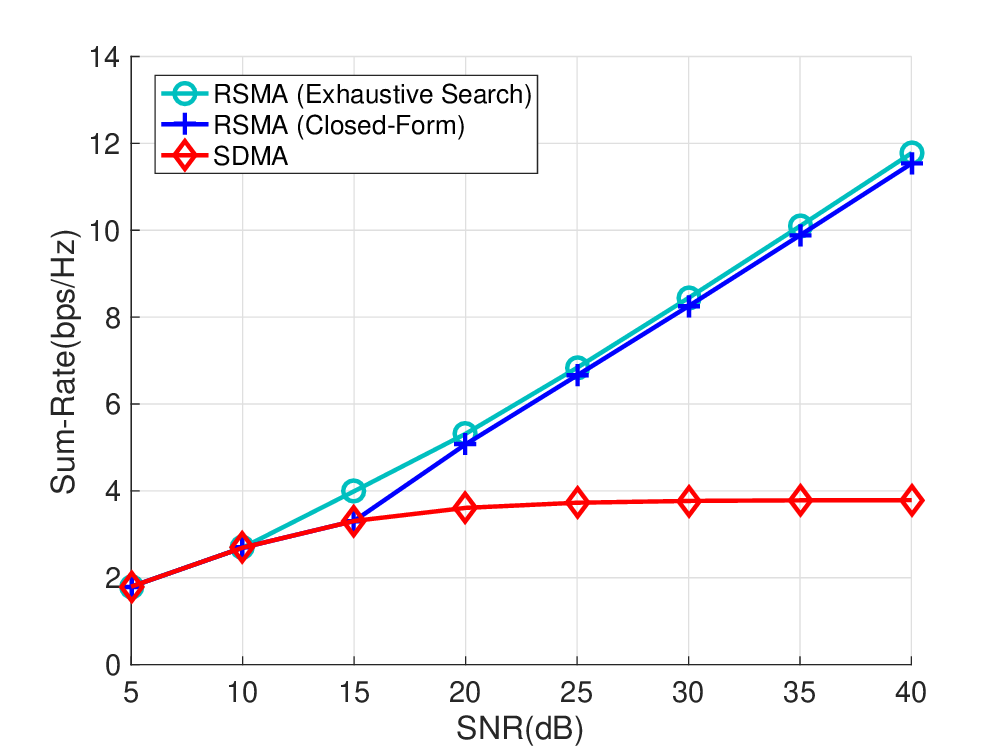}}
		\caption{$K=4$, $n_{t}=4$, Speed=$20$km/h.}
		\label{fig:sumrate_4420}
	\end{subfigure}
	\begin{subfigure}{.5\textwidth}
		\centerline{\includegraphics[width=3.1in,height=3.1in,keepaspectratio]{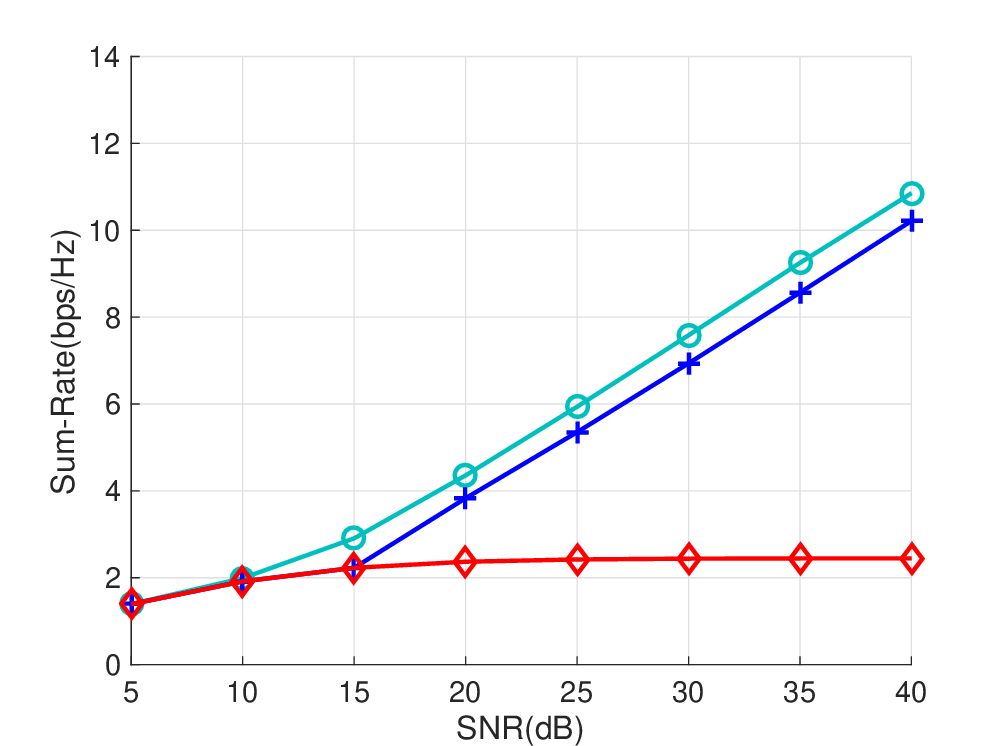}}
		\caption{$K=4$, $n_{t}=4$, Speed=$30$km/h.}
		\label{fig:sumrate_4430}
	\end{subfigure}
	\newline
	\begin{subfigure}{.5\textwidth}
		\centerline{\includegraphics[width=3.1in,height=3.1in,keepaspectratio]{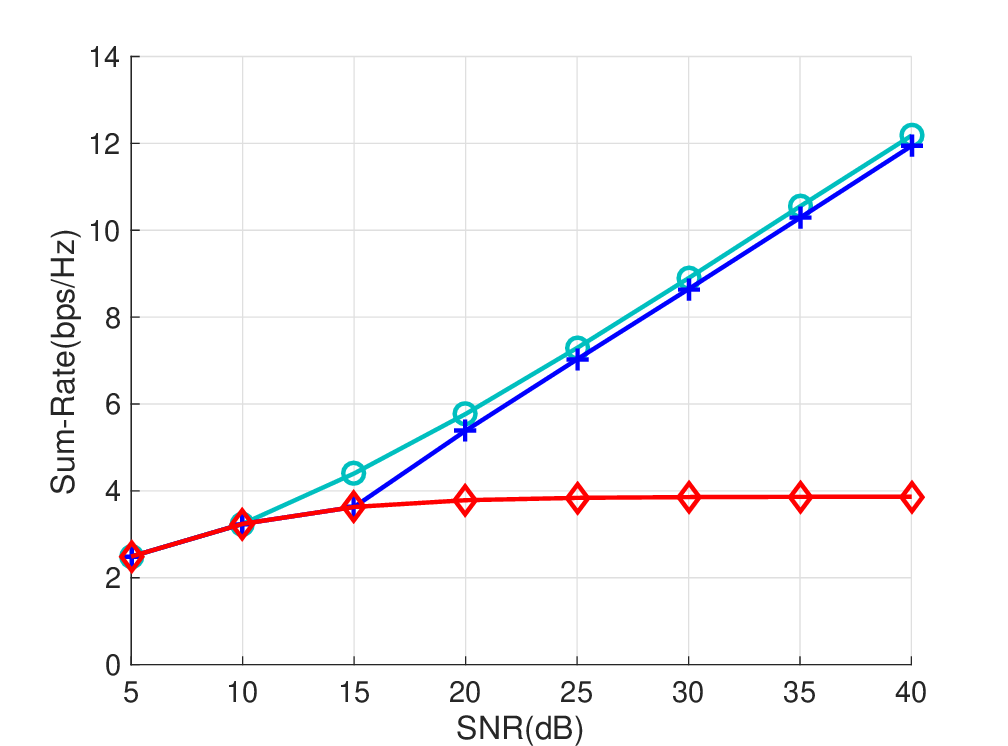}}
		\caption{$K=4$, $n_{t}=8$, Speed=$30$km/h.}
		\label{fig:sumrate_4830}
	\end{subfigure}
	\begin{subfigure}{.5\textwidth}
		\centerline{\includegraphics[width=3.1in,height=3.1in,keepaspectratio]{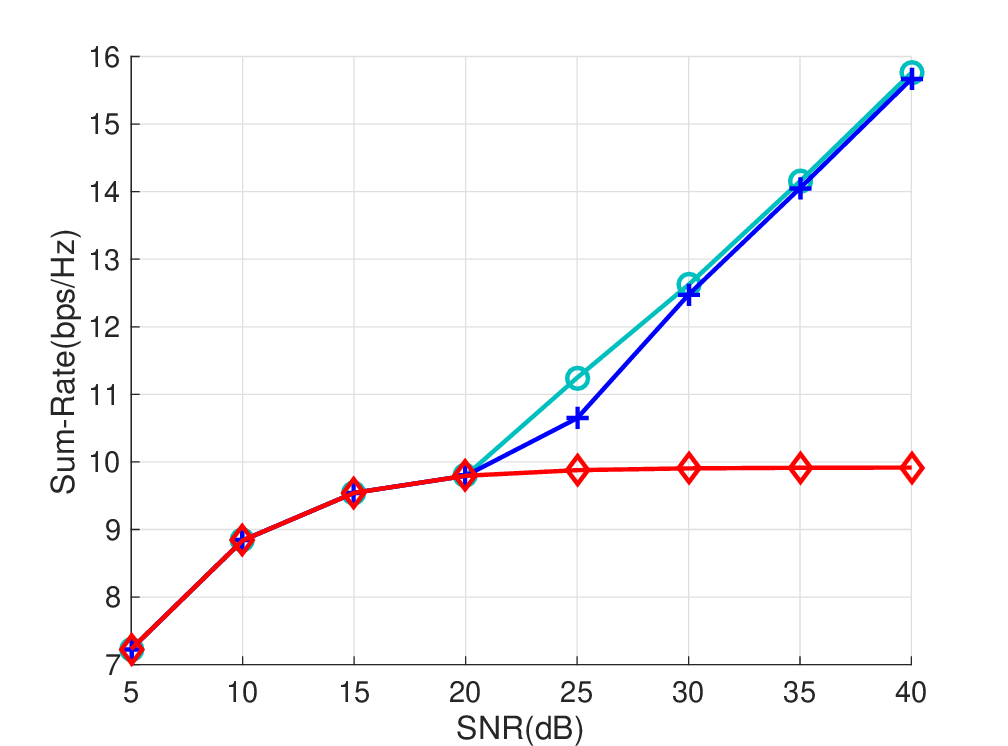}}
		\caption{$K=8$, $n_{t}=32$, user speed=$30$km/h.}
		\label{fig:sumrate_83230}
	\end{subfigure}
	\caption{Sum-rate vs. SNR.}
	\vspace{-0.5cm}
\end{figure*}

\section{Simulation Results}
\label{sec:simulation}
In this section, we demonstrate the behaviour of the lower bound compared to the ergodic sum-rate and the performance of 1-layer RSMA with the proposed power allocation algorithm by simulations. In the first part of the analysis, we perform Monte Carlo simulations over different channel realizations to investigate the sum-rate performance. 
We compare the sum-rate performance achieved by using the power allocation coefficient obtained from the expression \eqref{eq:poweralloc} and the coefficient that maximizes the lower bound \eqref{eqn:lowerbound}, which is found by exhaustive search. By such comparison, we aim to show that the proposed power allocation algorithm can achieve near-optimal performance. Furthermore, we compare the performance of SDMA and RSMA with the abovementioned power allocation methods to show that RSMA achieves significant performance gain over SDMA.   

In the second part of the analysis, we investigate the throughput performance by Link-Level Simulations (LLS) using the proposed power allocation algorithm with a practical transceiver architecture and a realistic 3rd Generation Partnership Project (3GPP) channel model. Moreover, we compare the performance of SDMA and RSMA with the proposed power allocation algorithm to verify the performance gain achieved by RSMA in a realistic setup.

\subsection{Sum-Rate Performance}
We start our analysis by investigating the behaviour of the derived lower bound in order to get an insight on the success of the obtained power allocation algorithm. 
Fig.~\ref{fig:lowerbound1}-\ref{fig:lowerbound8} show the ergodic sum-rate with Gaussian inputs obtained by expression \eqref{eqn:sumrate} and Monte Carlo simulations, and the lower bound calculated from expression \eqref{eqn:lowerbound}, both as functions of $t$ for different values of $\epsilon$, $n_{t}$, $K$, and $P$. We define SNR as $10\log_{10}(P)$, owing to the unit noise variance assumption in the system model (without loss of generality). We note that $t=1$ corresponds to the SDMA scheme, {\sl i.e.,} no power is allocated to the common stream and each message $W_{k}$ is directly encoded into a private stream.

An immediate observation from the figures is the improved sum-rate performance of RSMA ($0 \leq t\leq 1$) compared to SDMA ($t=1$), {\sl i.e.}, conventional MU-MIMO precoding. The optimal value of $t$ that maximizes the sum-rate of RSMA varies with the parameters $n_{t}$, $K$, $P$, $\epsilon$ and the maximal points at values $t < 1$ prove that message splitting is required to maximize the sum-rate. A general trend observed in all figures is that as the total transmit power increases, the increase in the sum-rate of SDMA saturates as a result of the multi-user interference becoming more dominant than noise (the system enters the interference-limited regime). Under such operating conditions, the optimal strategy to maximize the sum-rate is to allocate a certain portion of the total transmit power to the common stream. The figures show that the optimal portion varies with the system parameters and the CSIT quality. One can also note here that RSMA achieves significant performance gain over SDMA even with random beamformers employed for the common streams.

Another observation from Fig.~\ref{fig:lowerbound1}-\ref{fig:lowerbound8} is that the obtained lower bound captures the behaviour of the ergodic sum-rate with the varying parameters successfully. The results indicate that the power allocation coefficient $t$ that maximizes the obtained lower bound is almost identical to the coefficient that maximizes the actual ergodic sum-rate. Consequently, one can expect that the proposed low-complexity power allocation algorithm in \eqref{eq:poweralloc} achieves a near-optimal performance.

Next, we investigate the performance of 1-layer RSMA with the proposed power allocation algorithm and compare its performance with that of SDMA. We consider the ergodic sum-rate for Gaussian inputs. The performance of RSMA is studied by using power allocation coefficients obtained using two separate methods. For the first method, we perform an exhaustive search over the power allocation coefficient $t$ to find the value that maximizes the ergodic sum-rate expression given in \eqref{eqn:sumrate}. For the exhaustive search procedure, we consider all values of $t$ in the interval $\left(0,1 \right]$ with a granularity of $0.001$. For the second method, we use the coefficient $t_{opt}$ given by the closed-form solution in \eqref{eq:poweralloc}. We analyse the performance with the user speed values of $20$km/h and $30$km/h, a CSI acquisition delay of $2.5$ms and an operating frequency of $3.5$GHz. 
For an enhanced performance for the common stream, we consider the leftmost eigenvector (the eigenvector corresponding to the maximum eigenvalue) of the CSIT matrix \mbox{$\mathbf{H}[m-1]=\left[\mathbf{h}_{1}[m-1],\mathbf{h}_{1}[m-1],\ldots,\mathbf{h}_{K}[m-1] \right] $} as the common precoder. 
Fig.~\ref{fig:sumrate_4420}-\ref{fig:sumrate_83230}
show the ergodic sum-rates achieved by SDMA and RSMA with the two power allocation methods for different values of $K$, $n_{t}$, and user speed. 

As seen from the figures, the exhaustive search and the closed-form power allocation methods achieve similar sum-rate performance for all of the considered cases. The results demonstrate that the closed-form method provides a near-optimal solution for the ergodic sum-rate maximization problem with a low-complexity algorithm. 
An important conclusion from the figures is that RSMA achieves a non-saturating performance while the sum-rate of SDMA saturates in the interference limited SINR regime, as also expected from the results in Fig.~\ref{fig:lowerbound1}-\ref{fig:lowerbound8}. 
\begin{figure*}[t!]
	\begin{subfigure}{.5\textwidth}
		\centerline{\includegraphics[width=3.1in,height=3.1in,keepaspectratio]{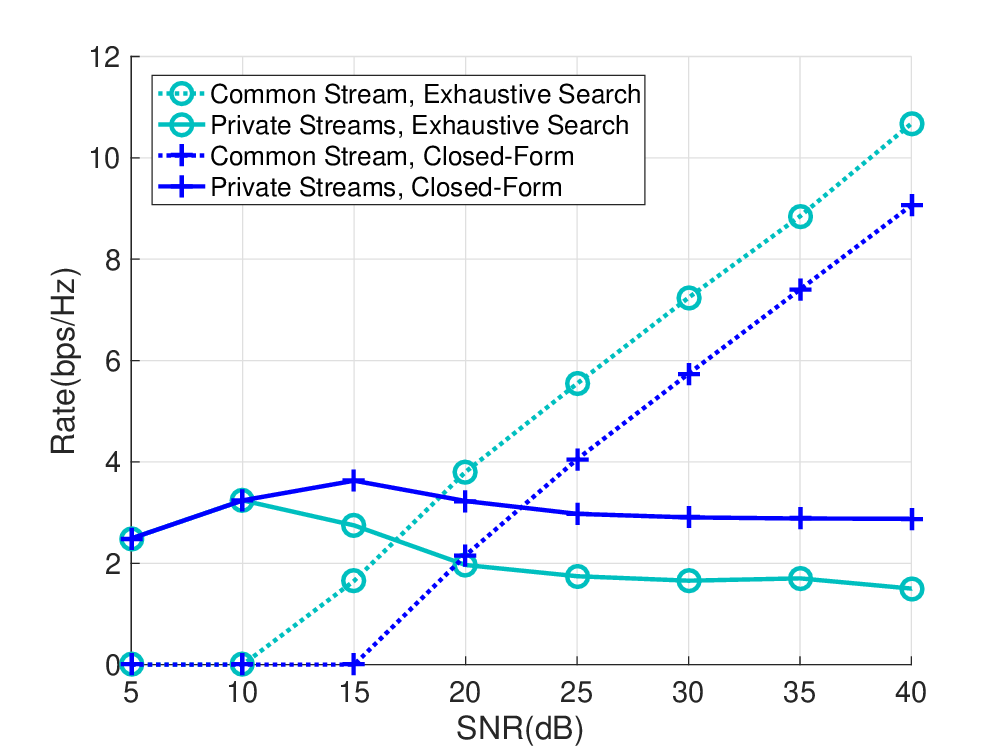}}
		\caption{$K=4$, $n_{t}=8$, user speed=$30$km/h.}
		\label{fig:streams_4830}
	\end{subfigure}
	\begin{subfigure}{.5\textwidth}
		\centerline{\includegraphics[width=3.1in,height=3.1in,keepaspectratio]{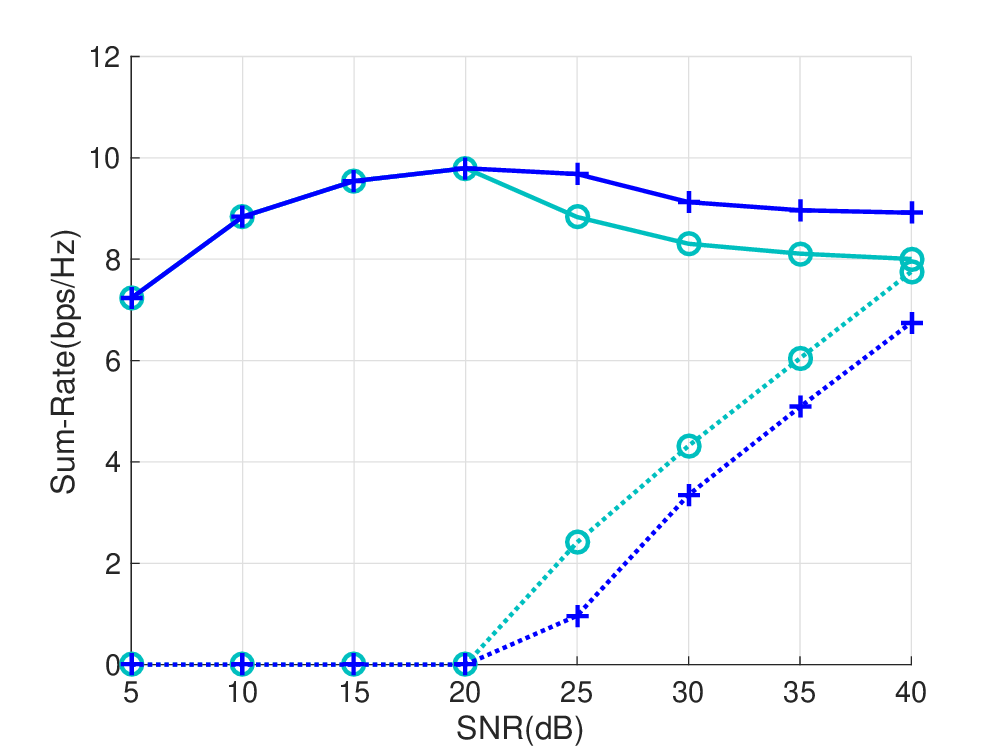}}
		\caption{$K=8$, $n_{t}=32$, user speed=$30$km/h.}
		\label{fig:streams_83230}
	\end{subfigure}
	\caption{Stream rates vs. user speed.}
	\vspace{-0.5cm}
\end{figure*}

Finally, we compare the behaviours of the closed-form and exhaustive search power allocation methods in different settings. For this purpose, we plot the rates allocated to the common and private streams separately.
Fig.~\ref{fig:streams_4830}-\ref{fig:streams_83230} show the rate allocations achieved by exhaustive search over the obtained lower bound and the closed-form solution for two different scenarios. 
The figures show that both methods behave similarly when distributing the power among the common and private streams. 
The portion of the total power allocated to the common stream increases with increasing SNR as the system moves into the interference-limited regime for both methods. Such behaviour allows RSMA to achieve a non-saturating sum-rate performance as opposed to SDMA, whose sum-rate saturates with SNR as observed from Fig.~\ref{fig:sumrate_4420}-\ref{fig:sumrate_83230}.	
The results imply that the proposed closed-form method achieves a near-optimal power allocation with a practical algorithm. 
\begin{figure*}[t!]
	\centerline{\includegraphics[width=6.0in,height=6.0in,keepaspectratio]{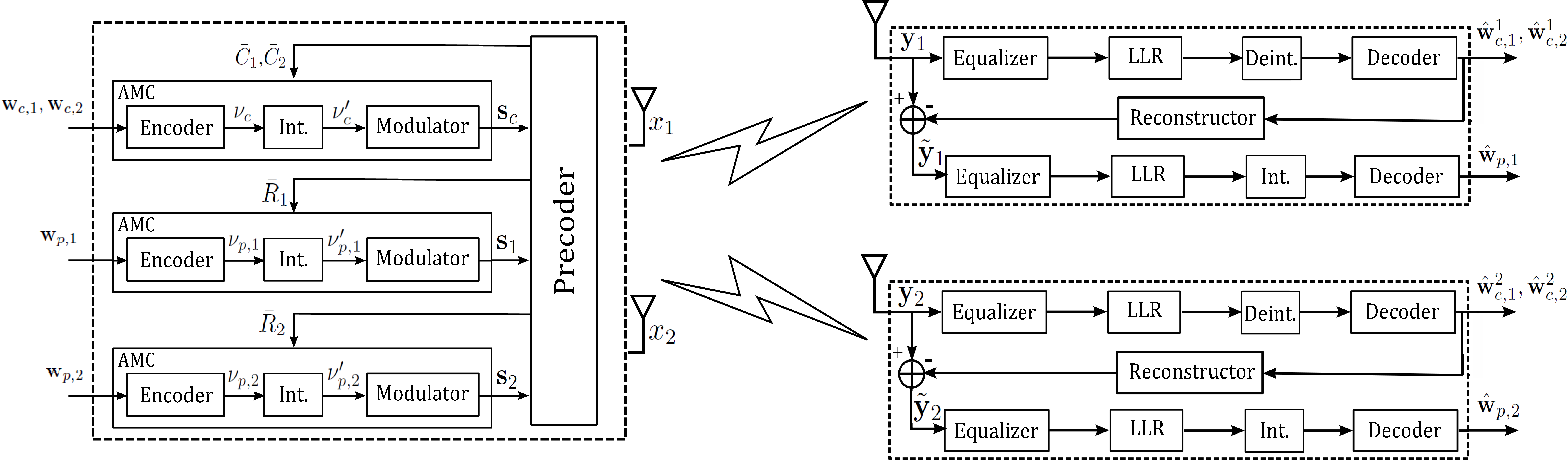}}
	\vspace{-0.15cm}
	\caption{Transmitter and receiver structures for 1-layer RSMA 
		\cite{dizdar_2020}.}
	\label{fig:transceiver}
	\vspace{-0.5cm}
\end{figure*}
\begin{figure*}[t!]
	\begin{subfigure}{.5\textwidth}
		\centerline{\includegraphics[width=3.1in,height=3.1in,keepaspectratio]{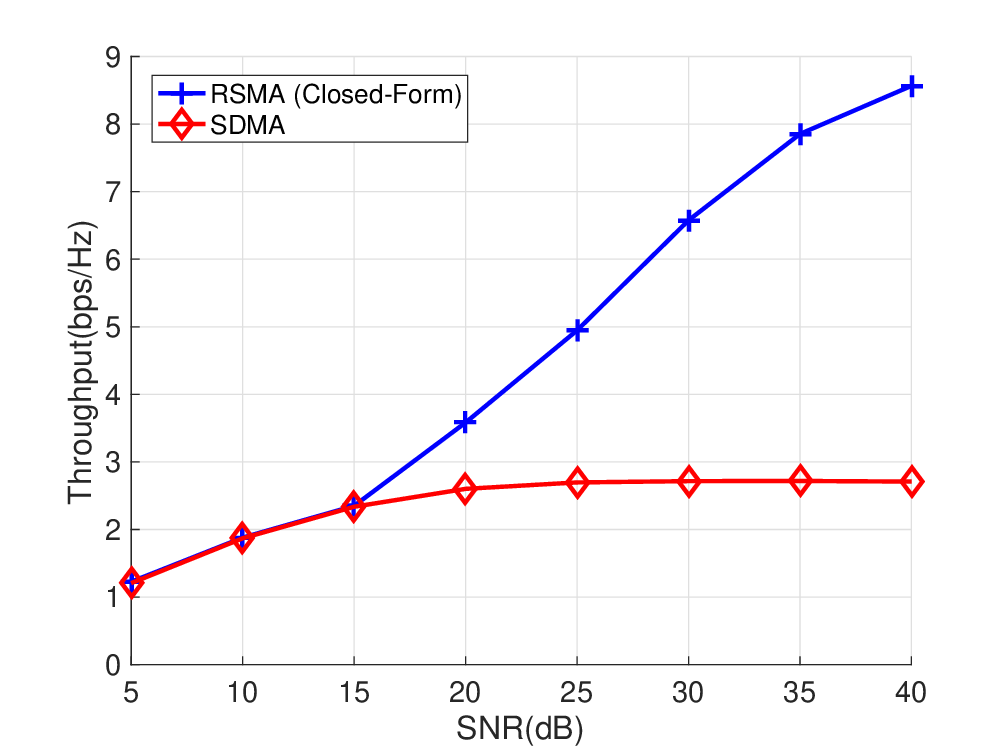}}
		\caption{$K=4$, $n_{t}=4$, Speed=$20$km/h.}
		\label{fig:tp_4420}
	\end{subfigure}
	\begin{subfigure}{.5\textwidth}
		\centerline{\includegraphics[width=3.1in,height=3.1in,keepaspectratio]{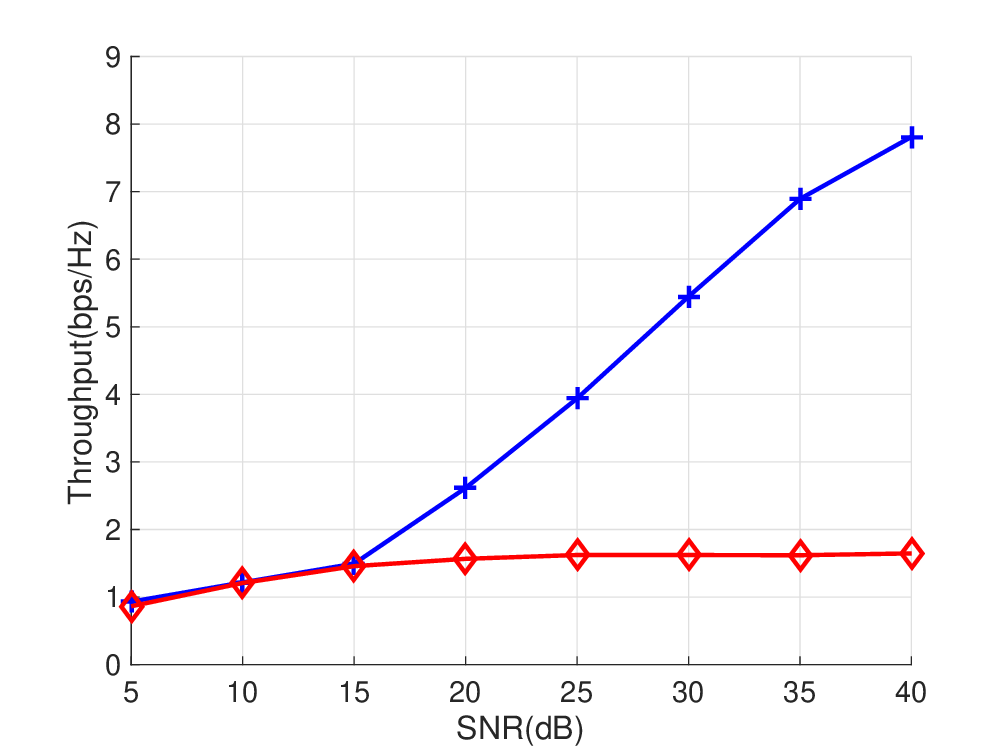}}
		\caption{$K=4$, $n_{t}=4$, Speed=$30$km/h.}
		\label{fig:tp_4430}
	\end{subfigure}
	\newline
	\begin{subfigure}{.5\textwidth}
		\centerline{\includegraphics[width=3.1in,height=3.1in,keepaspectratio]{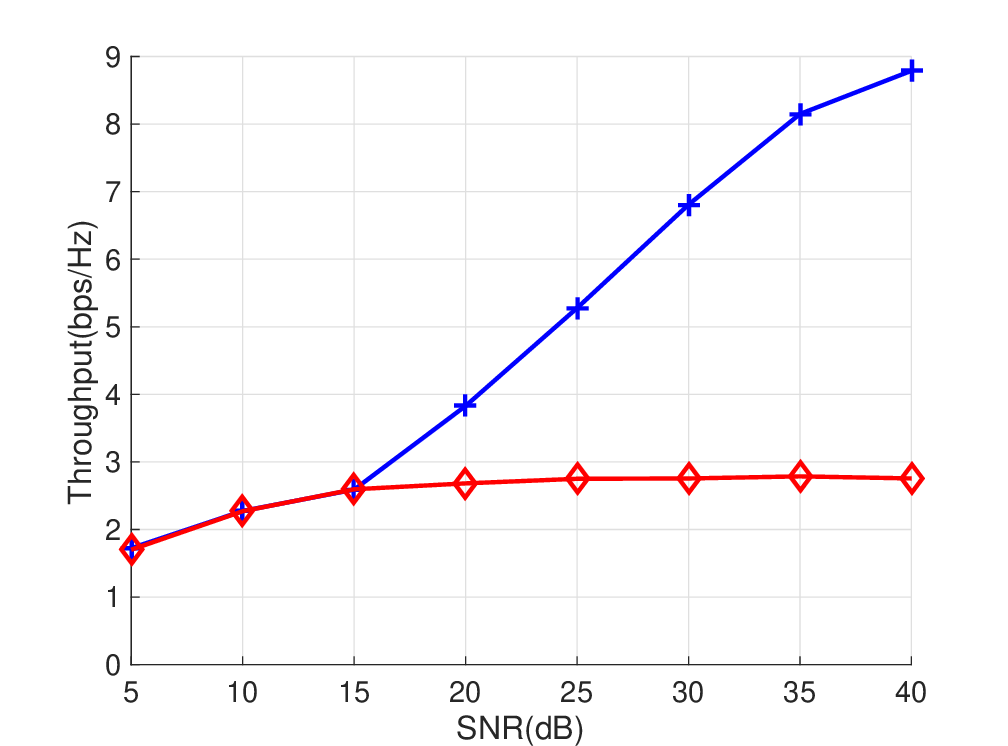}}
		\caption{$K=4$, $n_{t}=8$, Speed=$30$km/h.}
		\label{fig:tp_4830}
	\end{subfigure}
	\begin{subfigure}{.5\textwidth}
		\centerline{\includegraphics[width=3.1in,height=3.1in,keepaspectratio]{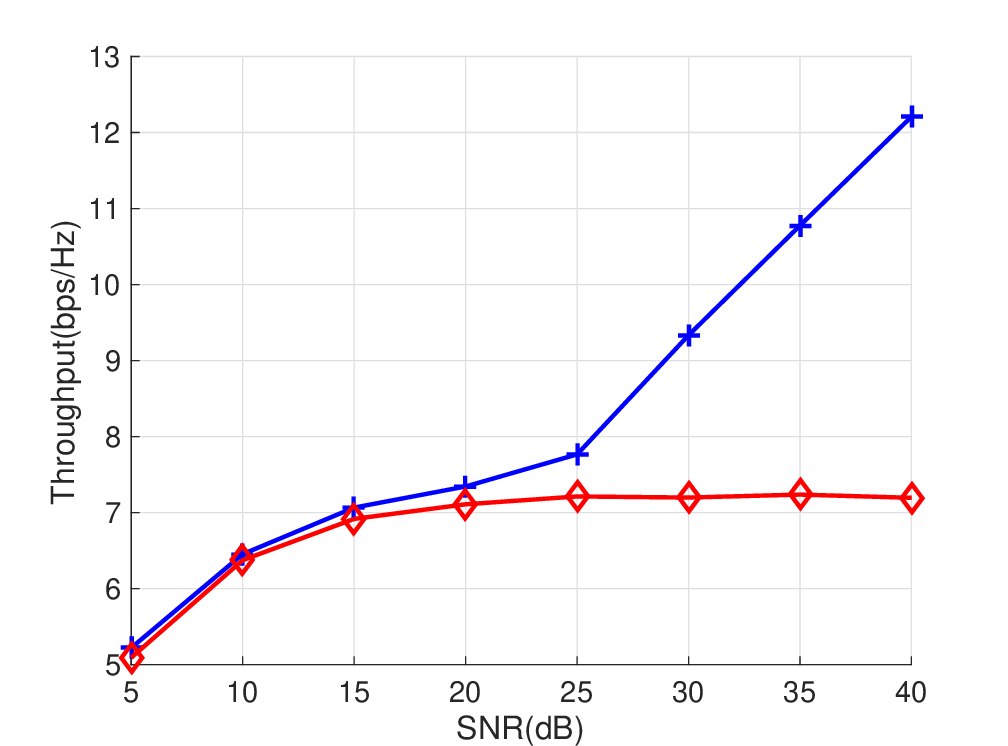}}
		\caption{$K=8$, $n_{t}=32$, Speed=$30$km/h.}
		\label{fig:tp_83230}
	\end{subfigure}
	\caption{Throughput vs. SNR.}
	\vspace{-0.5cm}
\end{figure*}

\subsection{Link-Level Simulations}
In this section, we investigate the throughput performance of SDMA and RSMA with the proposed power allocation algorithm by LLS. We use a transceiver architecture for RSMA with finite alphabet modulation schemes, finite-length channel coding, and an Adaptive Modulation and Coding (AMC) algorithm. We employ the modulation schemes $4$-QAM, $16$-QAM, $64$-QAM and $256$-QAM and polar coding \cite{arikan_2009}. 
The AMC algorithm selects a suitable modulation-coding rate pair for data transmission depending on the calculated transmit rates. The transmit rate calculations for AMC algorithm are performed assuming the instantaneous channel $\mathbf{h}_{k}[m]$, $\forall k \in \mathcal{K}$ are known in the AMC module (only for the purpose of modulation-coding scheme selection). 
For completeness, the overall architecture for two transmit antennas and two single-antenna users is presented in Fig.~\ref{fig:transceiver}, which is modified in this work to simulate a system with $K$ (single-antenna and multi-antenna) users.
The reader is referred to \cite{dizdar_2020} for more details on the architecture and the LLS platform.
\begin{figure*}[t!]
	\begin{subfigure}{.5\textwidth}
		\centerline{\includegraphics[width=3.1in,height=3.1in,keepaspectratio]{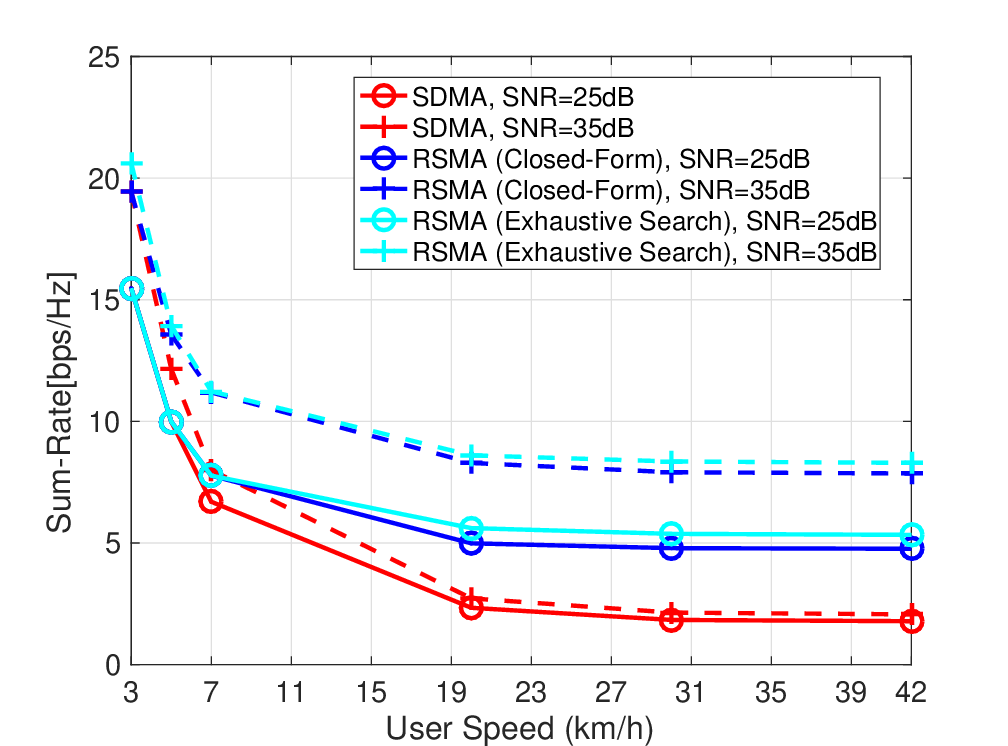}}
		\caption{Sum-rate vs. user speed.}
		\label{fig:cap_doppler_quadriga}
	\end{subfigure}
	\begin{subfigure}{.5\textwidth}
		\centerline{\includegraphics[width=3.1in,height=3.1in,keepaspectratio]{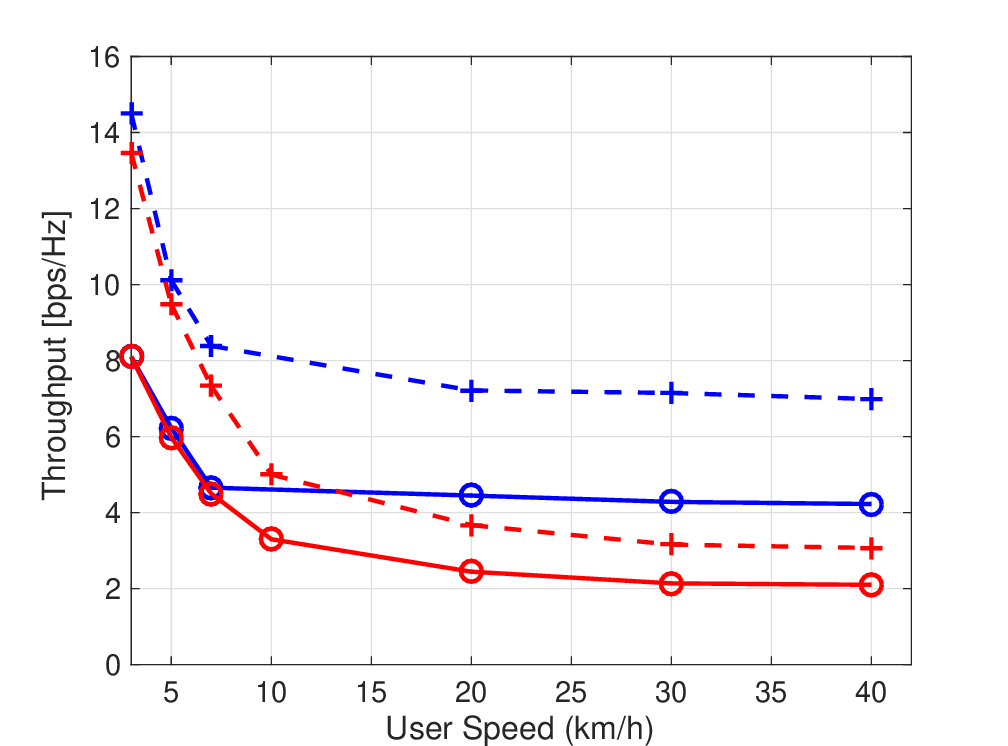}}
		\caption{Throughput vs. user speed.}
		\label{fig:tp_doppler_quadriga}
	\end{subfigure}
	\caption{Sum-rate and throughput vs. user speed, OFDM waveform, 3GPP channel model.}
	\vspace{-0.5cm}
\end{figure*}
\begin{figure*}[t!]
	\begin{subfigure}{.5\textwidth}
		\centerline{\includegraphics[width=3.1in,height=3.1in,keepaspectratio]{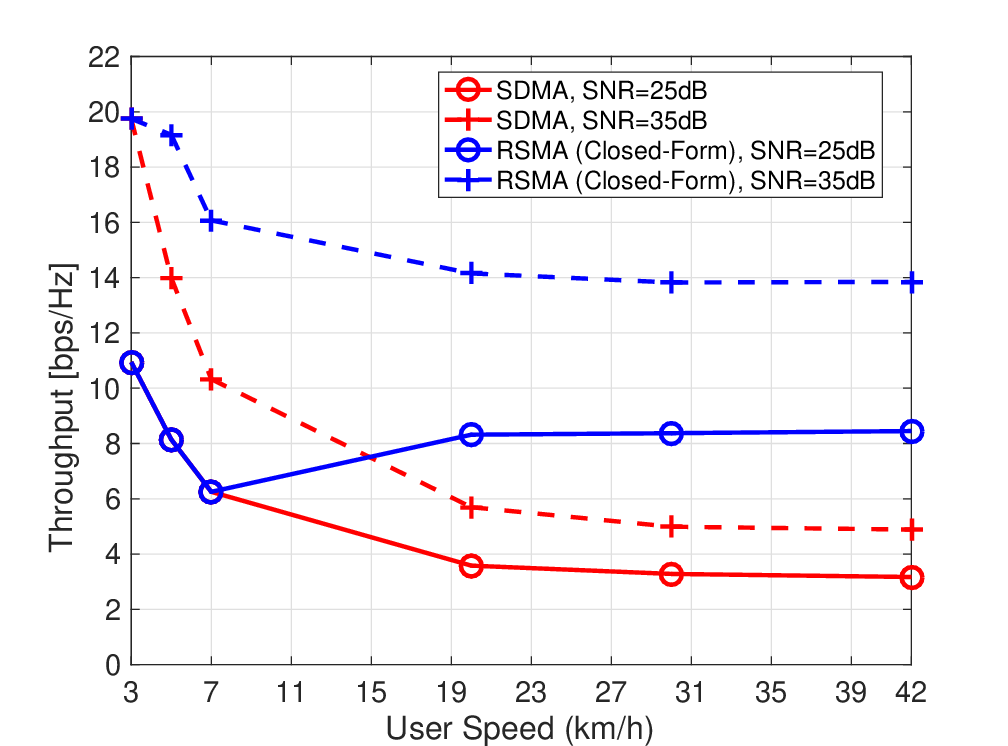}}
		\caption{Throughput vs. user speed, $Q_{c}=2$, $Q_{p}=2$.}
		\label{fig:tp_doppler_qc2qp2_quadriga}
	\end{subfigure}
	\begin{subfigure}{.5\textwidth}
		\centerline{\includegraphics[width=3.1in,height=3.1in,keepaspectratio]{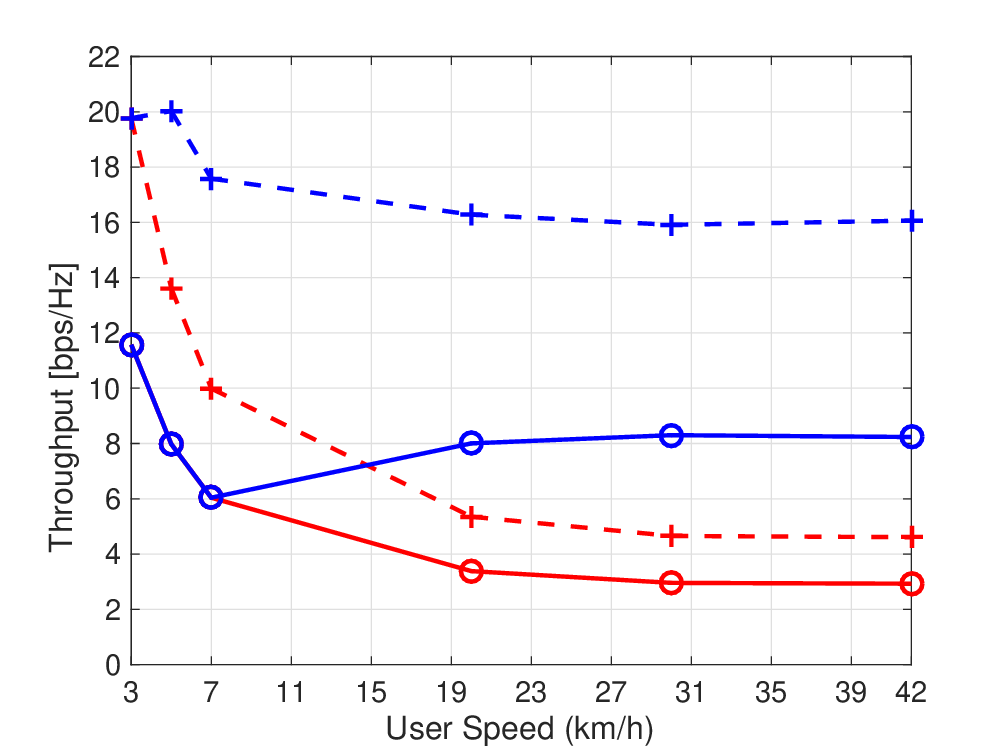}}
		\caption{Throughput vs. user speed, $Q_{c}=3$, $Q_{p}=3$.}
		\label{fig:tp_doppler_qc3qp3_quadriga}
	\end{subfigure}
	\caption{Sum-rate and throughput vs. user speed, OFDM waveform, 3GPP channel model, multiple receive antennas.}
	\vspace{-0.5cm}
\end{figure*}

Let $S^{(l)}$ denote the number of channel uses in the $l$-th Monte-Carlo realization and $D_{s,k}^{(l)}$ denote the number of successfully recovered information bits by user-$k$ in the common stream (excluding the common part of the message intended for the other user) and its private stream. Then, we calculate the throughput as  
\vspace{-0.3cm}
\begin{align}
	\mathrm{Throughput[bps/Hz]}=\frac{\sum_{l}(D_{s,1}^{(l)}+D_{s,2}^{(l)})}{\sum_{l}S^{(l)}}.
\end{align}

Fig.~\ref{fig:tp_4420}-\ref{fig:tp_83230} show the throughput performance of RSMA and SDMA for the settings used in Fig.~\ref{fig:sumrate_4420}-\ref{fig:sumrate_83230}. The results demonstrate that the throughput performance follow the exact same trends of the sum-rate curves. 
RSMA can achieve the performance enhancements demonstrated by the ergodic sum-rate analysis in terms of throughput with practical transmitter and receiver designs.

We take our performance analysis in practical scenarios one step further and study the performance of SDMA and RSMA with the proposed power allocation algorithm in more realistic channel models. We use the Quadriga Channel Generator \cite{jaeckel_2019} to generate channels according to the 3GPP Urban Macro-Cell channel model \cite{3gpp}. The channels of each user have a delay spread of $300$ns and $23$ clusters, with each cluster consisting of $20$ rays.

We use Cyclic-Prefix (CP)-OFDM waveform for transmission. The OFDM subcarrier spacing is set as $30$kHz. Data is carried over $256$ subcarriers. The CP length is $10\mu$s. We assume the system is operating at $3.5$GHz and the CSI acquisition delay is $10$ms. 
We calculate and apply precoders separately for each subcarrier. The power allocation coefficient $t_{opt}$ is constant on each subcarrier as the parameters $n_{t}$, $K$, $P$ and $\epsilon$ do not vary with subcarrier index. 

Fig.~\ref{fig:cap_doppler_quadriga}~and~\ref{fig:tp_doppler_quadriga} show the sum-rate and throughput performance of RSMA and SDMA with respect to user speed. We analyse the scenarios with $n_{t}=32$, $K=8$ and average received SNR levels of $25$dB and $35$dB in the simulations. 

First, we compare the performance of the two power allocation algorithms from Fig.~\ref{fig:cap_doppler_quadriga}. The power allocation coefficients obtained from the closed-form solution \eqref{eq:poweralloc} can achieve a performance close to the coefficients obtained by the exhaustive search for all user speed and SNR values, strengthening our conclusion from the previous results that the proposed power allocation algorithm can achieve near-optimal performance.  

Fig.~\ref{fig:cap_doppler_quadriga}~and~\ref{fig:tp_doppler_quadriga} clearly demonstrate the robustness of 1-layer RSMA with the proposed power allocation algorithm under user mobility. The sum-rate and achieved throughput of SDMA drops significantly with increasing user speed. Furthermore, the performance of SDMA saturates with increasing SNR, implying that the system operates in the interference-limited regime. On the other hand, RSMA limits the performance degradation starting at low user speed and achieves a robust performance as the speed increases. It can be observed that the sum-rate and achieved throughput of RSMA increases with increasing SNR, meaning that RSMA can operate in the interference-limited regime without a saturating throughput.

Although we consider single-antenna receivers in the derivations in this work, we extend our LLS analysis to a system model in which each receiver has $n_{r}$ receive antennas. We consider the transmission of a vector of $Q_{c}$ common streams to be decoded by all users and $K$ vectors of $Q_{p}$ private streams ({\sl i.e.,} the total number of streams is $Q_{c}+ KQ_{p}$). The precoders for the private streams are obtained by the Block-Diagonalization (BD) method to suppress the multi-user interference \cite{spencer_2004}.  
We employ the same power allocation algorithm in \eqref{eq:poweralloc} to distribute power among the precoders for the common and private streams. The power allocated to the common precoders is distributed equally among the common streams and the power allocated to the private precoders is distributed equally among the private streams of all users.

The receivers employ Minimum-Mean Square Error (MMSE) equalization to detect the intended common stream by suppressing the interference from the other common streams and all private streams. The common streams are decoded in parallel without any interference cancellation in-between. After all common streams are decoded, the overall common signal is reconstructed and interference cancellation is performed using the reconstructed common signal. Finally, the receivers apply an identical procedure to detect and decode the private streams.

Fig.~\ref{fig:tp_doppler_qc2qp2_quadriga}~and~\ref{fig:tp_doppler_qc3qp3_quadriga} demonstrate the throughput results for $n_{t}=64$, $n_{r}=4$, $K=8$ and different values of $Q_{c}$ and $Q_{p}$. The throughput curve of RSMA is observed to have a non-monotonic behaviour since the power allocation algorithm is sub-optimal for systems with multiple receive antennas, thus resulting in loss of throughput for specific user speed values. The figures clearly show that RSMA with the proposed power allocation algorithm can perform significantly better than SDMA also with multi-antenna receivers. Future works include the refinement of the lower bound derivations and the closed-form solution for the power allocation algorithm for systems with multi-antenna receivers.

\section{Conclusion}
\label{sec:conclusion}
In this work, we derive a lower bound for the approximate ergodic sum-rate of RSMA under user mobility for arbitrary number of transmit antennas, number of users, user speed and transmit power. Using this lower bound, we find a closed-form solution for the optimal power allocation coefficient that distributes the transmit power between the common stream and private streams. We perform simulations with realistic channel models to demonstrate the performance of RSMA with the obtained power allocation coefficient. Theoretical bounds and link-level simulation results clearly show that RSMA with the proposed power allocation scheme is robust to the degrading effects of user mobility and has significantly higher performance than that of SDMA. The key takeaway message is that in mobile MIMO networks, to maintain reliable multi-user connectivity, the transmission strategy should depart from conventional SDMA and Multi-User (massive) MIMO architecture that treats residual interference as noise, and adopts RSMA that is inherently robust to imperfect CSIT and mobility thanks to its ability to partially decode and partially treat interference as noise. 


%

\appendices
\section{Proof of Lemma 1}
\label{appendix:lemma1}
Using the expressions \eqref{eqn:dist1} and \eqref{eqn:dist2}, we can approximate the r.v. $X$ with the r.v. $\widehat{X}$, such that 
\begin{align}
	\widehat{X}= \epsilon^{2}|\mathbf{h}_{k}^{H}[m-1]\mathbf{p}_{k}|^{2}\hspace{-0.1cm}+(1-\epsilon^{2})\hspace{-0.1cm}\sum_{k\in\mathcal{K}}|\mathbf{e}_{k}^{H}[m]\mathbf{p}_{j}|^{2}.
	\label{eqn:gammasum}
\end{align}

As explained in Section~\ref{sec:lowerbound_private}, the terms $|\mathbf{h}_{k}^{H}[m-1]\mathbf{p}_{k}|^{2}$ and $|\mathbf{e}_{k}^{H}[m]\mathbf{p}_{j}|^{2}$ are r.v.s with Gamma distributions. Although the Gamma-distributed r.v.s in \eqref{eqn:gammasum} are correlated and have different scales, one can approximate $\widehat{X}$ with another Gamma-distributed random variable $\widetilde{X}$ by matching the first two moments \cite{jaramilloramirez_2015}.
Specifically, a random variable $Z=\sum_{i}A_{i}$, where $A_{i} \sim \mathrm{Gamma}(D_{i}, \theta_{i})$ can be approximated with a random variable $\widetilde{Z}\sim\mathrm{Gamma}(\widehat{D},\widehat{\theta})$, where
\begin{align}
	\widehat{D}=\frac{(\sum_{i}D_{i}\theta_{i})^{2}}{\sum_{i}D_{i}\theta_{i}^{2}}, \quad \quad 
	\widehat{\theta}=\frac{\sum_{i}D_{i}\theta_{i}^{2}}{\sum_{i}D_{i}\theta_{i}}. \label{eqn:gamma_parameters}
\end{align}
Therefore, we approximate $\widehat{X}$ by \mbox{$\widetilde{X} \sim \mathrm{Gamma}(\widehat{D},\widehat{\theta})$}, where 
\begin{align}
	&\widehat{D}=\frac{\left[ \epsilon^{2}(n_{t}+1)+(1-2\epsilon^{2})K\right] ^{2}}{\epsilon^{4}(n_{t}+1)+(1-2\epsilon^{2})K}, \nonumber \\ &\widehat{\theta}=\frac{\epsilon^{4}(n_{t}+1)+(1-2\epsilon^{2})K}{\epsilon^{2}(n_{t}+1)+(1-2\epsilon^{2})K}. \nonumber
\end{align}

\section{Proof of Lemma 2}
\label{appendix:lemma2}
We start the proof by finding the CDF of the r.v. $Y_{k}$.
\begin{align}
	Y_{k}=\frac{Q_{k1}}{1+\frac{P\widehat{\theta}t}{K}Q_{k2}},
\end{align}
with \mbox{$Q_{k1}=|\mathbf{h}^{H}_{k}[m]\mathbf{p}_{c}|^{2}$} and
$
Q_{k2}=\frac{1}{\widehat{\theta}}\sum_{j\in\mathcal{K}} |\mathbf{h}^{H}_{k}[m]\mathbf{p}_{j}|^{2}$.

It is assumed that the precoder for the common stream,  $\mathbf{p}_{c}$, to be independent of $\mathbf{h}_{k}[m]$, $\forall k\in \mathcal{K}$ and isotropically distributed on the unit sphere \cite{hao_2015}. As discussed in Section~\ref{sec:lowerbound_private}, these assumptions result in \mbox{$|\mathbf{h}^{H}_{k}[m]\mathbf{p}_{c}|^{2}\sim \mathrm{Gamma}(1,1)$}.

Lemma 1 states that the r.v. $X=\sum_{j\in\mathcal{K}} |\mathbf{h}^{H}_{k}[m]\mathbf{p}_{j}|^{2}$ can be approximated by another r.v. \mbox{$\widetilde{X} \sim \mathrm{Gamma}(\widehat{D},\widehat{\theta})$} with $\widehat{D}$ and $\widehat{\theta}$ given in \eqref{eqn:dandtheta}. 
Using this result, we can approximate $Q_{k2}$ with a r.v. \mbox{$\widetilde{Q}_{k2} \sim \mathrm{Gamma}(\widehat{D},1)$}. 

For tractability, we assume independence among the terms in the numerator and the denominator. Then, we can approximate $Y_{k}$ by $\widetilde{Y}_{k}$ with the CDF
\begin{align}
	F_{\widetilde{Y}_{k}}(y)&=\int_{0}^{\infty}P\left({Q}_{k1}<y\left( 1+\frac{P\widehat{\theta}t}{K}q\right) \right) f_{\widetilde{Q}_{k2}}(q)dq \nonumber\\
	&=\int_{0}^{\infty}\left(1-e^{-y\left(1+ \frac{P\widehat{\theta}t}{K}q\right) } \right) \frac{e^{-q}q^{\widehat{D}-1}}{\Gamma(\widehat{D})}dq \nonumber\\
	&=\int_{0}^{\infty}\frac{e^{-q}q^{\widehat{D}-1}}{\Gamma(\widehat{D})}dq -\frac{e^{-y}}{\Gamma(\widehat{D})}\int_{0}^{\infty}\hspace{-0.3cm}e^{-q\left(y\frac{P\widehat{\theta}t}{K}+1 \right) }q^{\widehat{D}-1}dq \nonumber \\
	&=1 -\frac{e^{-y}}{\Gamma(\widehat{D})}\int_{0}^{\infty}e^{-q\left(y\frac{P\widehat{\theta}t}{K}+1 \right) }q^{\widehat{D}-1}dq.
	\label{eqn:cdf_1}
\end{align} 
We perform a change of variables by $z=q\left(y\frac{P\widehat{\theta}t}{K}+1 \right)$, so that we can rewrite \eqref{eqn:cdf_1} as
\begin{align}
	&F_{\widetilde{Y}_{k}}(y) \nonumber \\
	&=1-\frac{e^{-y}}{\Gamma(\widehat{D})}\int_{0}^{\infty}e^{-z}\frac{z^{\widehat{D}-1}}{\left(y\frac{P\widehat{\theta}t}{K}+1 \right)^{(\widehat{D}-1)} }\frac{dz}{\left(y\frac{P\widehat{\theta}t}{K}+1 \right)}, \nonumber\\
	&=1 -\frac{e^{-y}}{\left(y\frac{P\widehat{\theta}t}{K}+1 \right)^{\widehat{D}}}\int_{0}^{\infty}\frac{e^{-z}z^{\widehat{D}-1}}{\Gamma(\widehat{D})}dz \nonumber \\
	&=1 -\frac{e^{-y}}{\left(y\frac{P\widehat{\theta}t}{K}+1 \right)^{\widehat{D}}}, 
\end{align} 
with the support set $[0,\infty)$.

Using the assumption that $Y_{k}$ are independent \cite{hao_2015}, we can approximate $Y$ by $\widetilde{Y}$ with the CDF \mbox{$F_{\widetilde{Y}}(y)=(1-(1-F_{\widetilde{Y}_{k}}(y))^{K})$}, which yields
\begin{align}
	F_{\widehat{Y}}(y)&=1-\left(1-1 + \frac{e^{-y}}{\left(y\frac{P\widehat{\theta}t}{K}+1 \right)^{\widehat{D}} } \right)^{K} \nonumber\\
	&=1-\frac{e^{-Ky}}{\left(y\frac{P\widehat{\theta}t}{K}+1 \right)^{\widehat{D}K} }.
	\label{eqn:cdf_minY}
\end{align}

\section{Proof of Lemma 4}
\label{appendix:lemma4}
We focus on the term $e^{\frac{K^{2}}{P\widehat{\theta}t}}\sum_{m=1}^{\left \lfloor{\widehat{D}K}\right \rceil}\mathrm{E}_{m}\left(\frac{K^{2}}{P\widehat{\theta}t}\right)$ since it is the only term in $\beta$ that is a function of $P$ and $t$.
In order to approximate this term under the assumption $Pt\rightarrow \infty$, we make use of a recurrence relation given in \cite{chiccoli_1990} as 
\begin{align}
	m\mathrm{E}_{m+1}(x)=e^{-x}-x\mathrm{E}_{m}(x).
	\label{eqn:recursive_exp}
\end{align}

Starting with the relation \eqref{eqn:recursive_exp}, one can write
\begin{align}
	\mathrm{E}_{m}(x)\hspace{-0.1cm}=\hspace{-0.1cm}\frac{e^{-x}}{m-1}\hspace{-0.1cm}+\hspace{-0.1cm}\frac{e^{-x}}{(m-1)!}\hspace{-0.1cm}\sum_{k=1}^{m-2}\hspace{-0.05cm}(-x)^{k}\hspace{-0.1cm}+\hspace{-0.1cm}\frac{(-x)^{m-1}}{(m-1)!}\mathrm{E}_{1}(x),\hspace{-0.05cm}
	\label{eqn:em_e1}
\end{align}
for any integer $m>1$. 

We use \eqref{eqn:em_e1} to expand $e^{\frac{K^{2}}{P\widehat{\theta}t}}\sum_{m=1}^{\left \lfloor{\widehat{D}K}\right \rceil}\mathrm{E}_{m}\left(\frac{K^{2}}{P\widehat{\theta}t}\right)$. The resulting expression is given in \eqref{eqn:beta_approx4}. 
\begin{table*}[t!]
	\begin{subequations}
		\begin{align}
			&e^{\frac{K^{2}}{P\widehat{\theta}t}}\sum_{m=1}^{\left \lfloor{\widehat{D}K}\right \rceil}\mathrm{E}_{m}\left(\frac{K^{2}}{P\widehat{\theta}t}\right)\nonumber \\
			&=e^{\frac{K^{2}}{P\widehat{\theta}t}}\left( \mathrm{E}_{1}\left( \frac{K^{2}}{P\widehat{\theta}t}\right) 
			+\sum_{m=2}^{\left \lfloor{\widehat{D}K}\right \rceil}\frac{e^{-\frac{K^{2}}{P\widehat{\theta}t}}}{m-1} +\sum_{m=3}^{\floor{\widehat{D}K}}\frac{e^{-\frac{K^{2}}{P\widehat{\theta}t}}}{(m-1)!}\sum_{k=1}^{m-2}\left( -\frac{K^{2}}{P\widehat{\theta}t}\right)^{k}
			+\mathrm{E}_{1}\left( \frac{K^{2}}{P\widehat{\theta}t}\right) \sum_{m=2}^{\left \lfloor{\widehat{D}K}\right \rceil}\frac{(-\frac{K^{2}}{P\widehat{\theta}t})^{m-1}}{(m-1)!}\right)\label{eqn:beta_approx1}   \\
			&\approx \gamma+\ln\left( \left \lfloor{\widehat{D}K}\right \rceil-1\right) +\frac{1}{2\left( \left \lfloor{\widehat{D}K}\right \rceil-1\right) }
			+\sum_{m^{\prime}=2}^{\left \lfloor{\widehat{D}K}\right \rceil-1}\frac{1}{m^{\prime}!}\sum_{k=1}^{m^{\prime}-1}\left( -\frac{K^{2}}{P\widehat{\theta}t}\right)^{k}
			+e^{\frac{K^{2}}{P\widehat{\theta}t}}\mathrm{E}_{1}\left( \frac{K^{2}}{P\widehat{\theta}t}\right) \left( 1+\hspace{-0.3cm}\sum_{m^{\prime}=1}^{\left \lfloor{\widehat{D}K}\right \rceil-1}\hspace{-0.1cm}\frac{\left( -\frac{K^{2}}{P\widehat{\theta}t}\right) ^{m^{\prime}}}{m^{\prime}!} \right)\label{eqn:beta_approx2}   \\
			&=\gamma+\ln\left( \left \lfloor{\widehat{D}K}\right \rceil-1\right) +\frac{1}{2\left( \left \lfloor{\widehat{D}K}\right \rceil-1\right) }+\sum_{m^{\prime}=2}^{\left \lfloor{\widehat{D}K}\right \rceil-1}\hspace{-0.1 cm}\frac{-\frac{K^{2}}{P\widehat{\theta}t}-\left( -\frac{K^{2}}{P\widehat{\theta}t}\right) ^{m^{\prime}}}{m^{\prime}!\left( 1+\frac{K^{2}}{P\widehat{\theta}t}\right) }+e^{\frac{K^{2}}{P\widehat{\theta}t}}\mathrm{E}_{1}\left( \frac{K^{2}}{P\widehat{\theta}t}\right) \left( 1+\hspace{-0.3cm}\sum_{m^{\prime}=1}^{\left \lfloor{\widehat{D}K}\right \rceil-1}\hspace{-0.1cm}\frac{\left( -\frac{K^{2}}{P\widehat{\theta}t}\right) ^{m^{\prime}}}{m^{\prime}!} \right)\label{eqn:beta_approx3}  \\
			&=\gamma+\ln\left( \left \lfloor{\widehat{D}K}\right \rceil-1\right) +\frac{1}{2\left( \left \lfloor{\widehat{D}K}\right \rceil-1\right) }+\phi(K,P,t,\widehat{D},\widehat{\theta}).
			\label{eqn:beta_approx4}
		\end{align} 
		\label{eqn:beta_approx}
	\end{subequations}
	\hrule
	\vspace{-0.8cm}
\end{table*}
The term $\phi(K,P,t,\widehat{D},\widehat{\theta})$ in \eqref{eqn:beta_approx4} is defined as
\begin{align}
	&\phi(K,P,t,\widehat{D},\widehat{\theta})\triangleq e^{\frac{K^{2}}{P\widehat{\theta}t}}\mathrm{E}_{1}\left( \frac{K^{2}}{P\widehat{\theta}t}\right) \left( \frac{\Gamma\left( \left \lfloor{\widehat{D}K}\right \rceil-1,-\frac{K^{2}}{P\widehat{\theta}t}\right) }{e^{\frac{K^{2}}{P\widehat{\theta}t}}\Gamma\left( \left \lfloor{\widehat{D}K}\right \rceil-1\right) } \right)\nonumber \\
	&-\frac{\frac{K^{2}}{P\widehat{\theta}t}}{\left( 1+\frac{K^{2}}{P\widehat{\theta}t}\right) }\left( \frac{\floor{\left( \left \lfloor{\widehat{D}K}\right \rceil-1\right) !e}}{\left( \left \lfloor{\widehat{D}K}\right \rceil-1\right) !}-2\right)\nonumber \\
	&-\frac{1}{\left( 1+\frac{K^{2}}{P\widehat{\theta}t}\right) }\left( \frac{\Gamma\left( \left \lfloor{\widehat{D}K}\right \rceil-1,-\frac{K^{2}}{P\widehat{\theta}t}\right) }{e^{\frac{K^{2}}{P\widehat{\theta}t}}\Gamma\left( \left \lfloor{\widehat{D}K}\right \rceil-1\right) }-1+\frac{K^{2}}{P\widehat{\theta}t}\right)\hspace{-0.1cm}.
	\label{eqn:phi}
\end{align}
The approximation in \eqref{eqn:beta_approx2}
follows from \mbox{$\sum^{n}_{k=1}\frac{1}{k}\approx\ln n + \gamma + \frac{1}{2n}$} for the finite harmonic series, applied on the second term in \eqref{eqn:beta_approx1}, such that
\begin{align}
	e^{\frac{K^{2}}{P\widehat{\theta}t}}\hspace{-0.1cm}\sum_{m=2}^{\left \lfloor{\widehat{D}K}\right \rceil}\hspace{-0.1cm}\frac{e^{-\frac{K^{2}}{P\widehat{\theta}t}}}{m-1} &=\sum_{m^{\prime}=1}^{\left \lfloor{\widehat{D}K}\right \rceil-1}\frac{1}{m^{\prime}} \nonumber \\
	&=\gamma+\ln\left( \left \lfloor{\widehat{D}K}\right \rceil-1\right) +\frac{1}{2\left( \left \lfloor{\widehat{D}K}\right \rceil-1\right)}. \nonumber
\end{align}
The expression \eqref{eqn:beta_approx3} follows from the solution for the alternating geometric series, i.e.,
\begin{align}
	\sum_{k=1}^{m-2}\left( -\frac{K^{2}}{P\widehat{\theta}t}\right) ^{k}=
	\frac{-\frac{K^{2}}{P\widehat{\theta}t}-\left( -\frac{K^{2}}{P\widehat{\theta}t}\right) ^{m}}{\left( 1+\frac{K^{2}}{P\widehat{\theta}t}\right) }.
\end{align}
The second and third terms in \eqref{eqn:phi} follow from the properties
\begin{align}
	\sum^{n}_{k=0}\frac{1}{k!} =\frac{\left \lfloor{n!e}\right \rfloor }{n!}, \quad \sum^{n}_{k=0}\frac{(-x)^{k}}{k!} =\frac{\Gamma(n,-x)}{e^{x}\Gamma(n)},
\end{align}   
applied to the last two terms in \eqref{eqn:beta_approx3}.

By the assumptions $Pt\rightarrow\infty$ and $e^{x}\mathrm{E}_{1}(x)\approx -\gamma - \ln(x)$ as $x \rightarrow 0$, we approximate \eqref{eqn:beta_approx4} as 
\begin{align}
	&e^{\frac{K^{2}}{P\widehat{\theta}t}}\hspace{-0.1cm}\sum_{m=1}^{\left \lfloor{\widehat{D}K}\right \rceil}\hspace{-0.1cm}\mathrm{E}_{m}\left(\frac{K^{2}}{P\widehat{\theta}t}\right)\nonumber \\
	&\quad \approx\hspace{-0.05cm}\ln\left( \left \lfloor{\widehat{D}K}\right \rceil\hspace{-0.1cm}-\hspace{-0.1cm}1\right)\hspace{-0.1cm}+\hspace{-0.1cm}\frac{1}{2\left( \left \lfloor{\widehat{D}K}\right \rceil\hspace{-0.1cm}-\hspace{-0.1cm}1\hspace{-0.05cm}\right) }\hspace{-0.05cm}-\hspace{-0.1cm}\ln\hspace{-0.05cm}\left(\hspace{-0.05cm}\frac{K^{2}}{P\widehat{\theta}t} \right).
	\label{eqn:prefinal_betaapprox}
\end{align}
Finally, substituting \eqref{eqn:prefinal_betaapprox} into the expression for $\beta$, we get \eqref{eqn:beta_approx0}.  
\vspace{-0.6cm}



\ifCLASSOPTIONcaptionsoff
  \newpage
\fi



%

%

\begin{IEEEbiography}[{\includegraphics[width=1in,height=1.5in,clip,keepaspectratio]{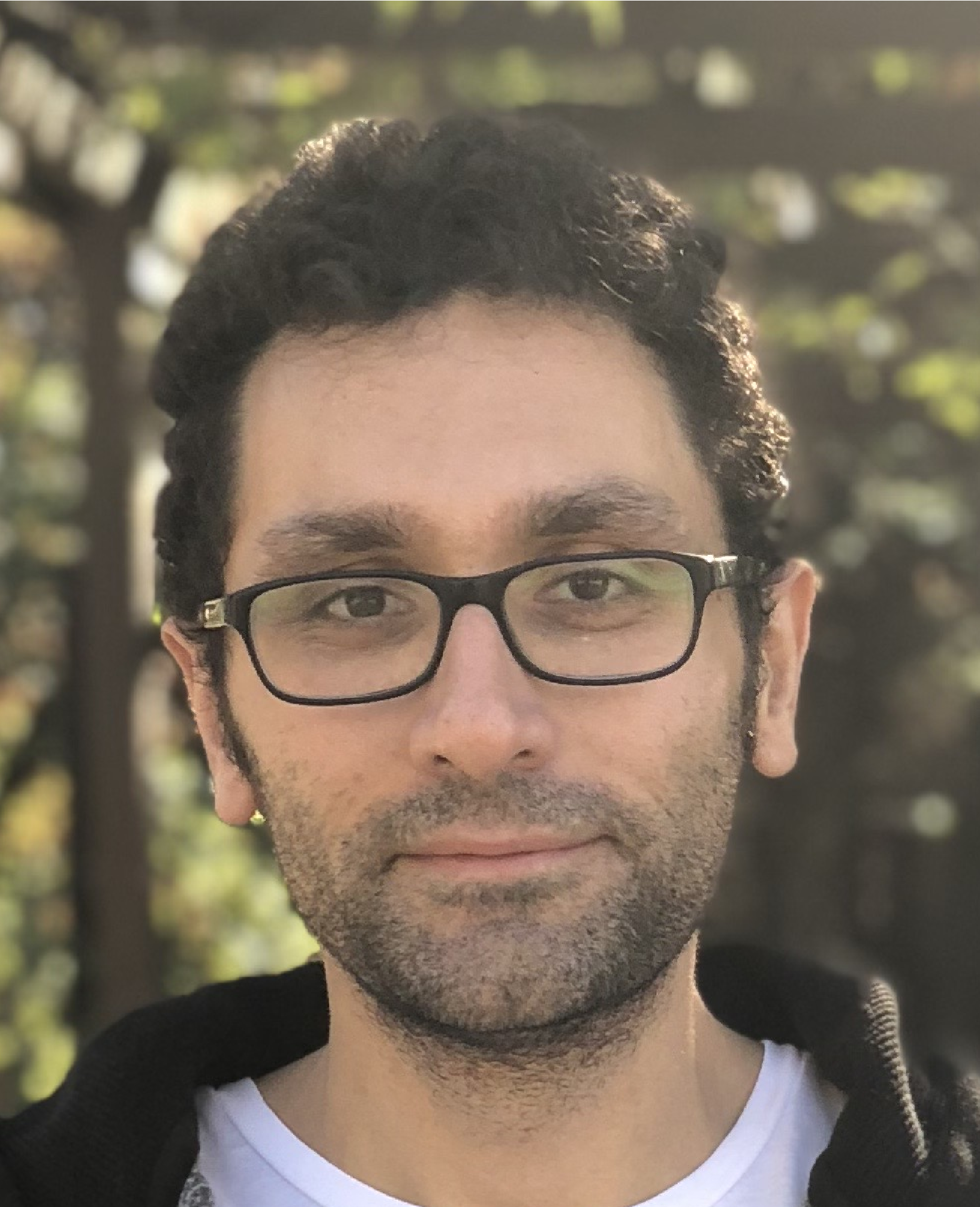}}]{Onur Dizdar}
Onur Dizdar received the B.Sc. and M.Sc. degrees in Electrical and Electronics Engineering from Middle East Technical University, Ankara, Turkey, in 2008 and 2011. He received his Ph.D. in Electrical and Electronics Engineering from Bilkent University, Ankara, Turkey, in 2017. He also worked as a communications system design engineer in ASELSAN, Turkey from 2008 to 2019. He is currently a postdoctoral research associate in the Communications and Signal Processing Group at Imperial College London. His research interests include wireless communications, error-correcting codes and decoding algorithms, rate-splitting and non-orthogonal multiple access and signal processing. 
\end{IEEEbiography}

\begin{IEEEbiography}[{\includegraphics[width=1in,height=1.25in,clip,keepaspectratio]{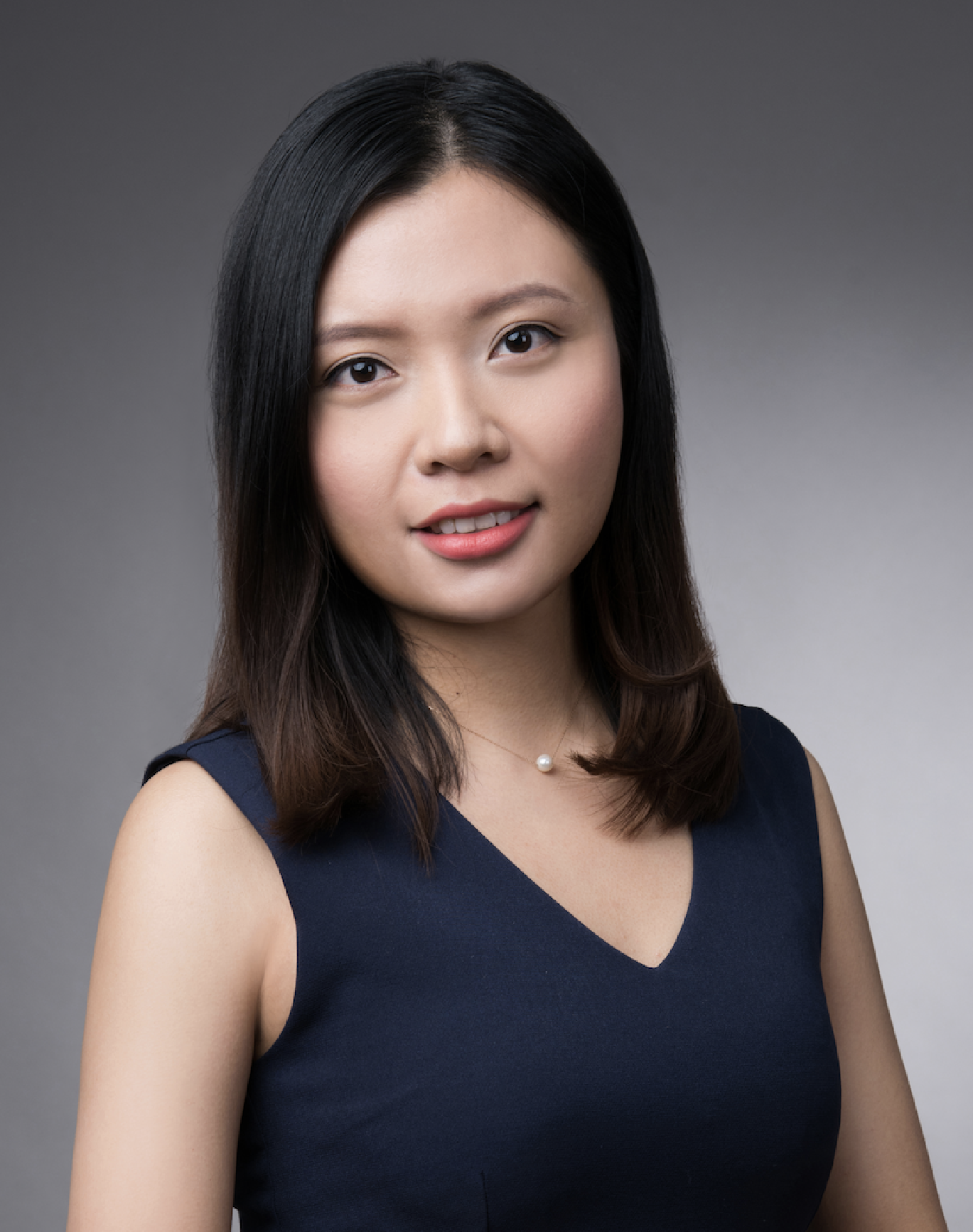}}]{Yijie Mao}
Yijie Mao received the B.Eng. degree from Beijing University of Posts and Telecommunications, and the B.Eng. (Hons.) degree from Queen Mary University of London (London, United Kingdom) in 2014. She received the Ph.D. degree in the Electrical and Electronic Engineering Department from the University of Hong Kong (Hong Kong, China) in 2018. She was a Postdoctoral Research Fellow at the University of Hong Kong from 2018 to 2019. She is currently a postdoctoral research associate with the Communications and Signal Processing Group (CSP), Department of the Electrical and Electronic Engineering at Imperial College London. Her research interests include Multiple Input Multiple Output (MIMO) communication networks, rate-splitting and non-orthogonal multiple access for 5G and beyond.
\end{IEEEbiography}


\begin{IEEEbiography}[{\includegraphics[width=1in,height=1.25in,clip,keepaspectratio]{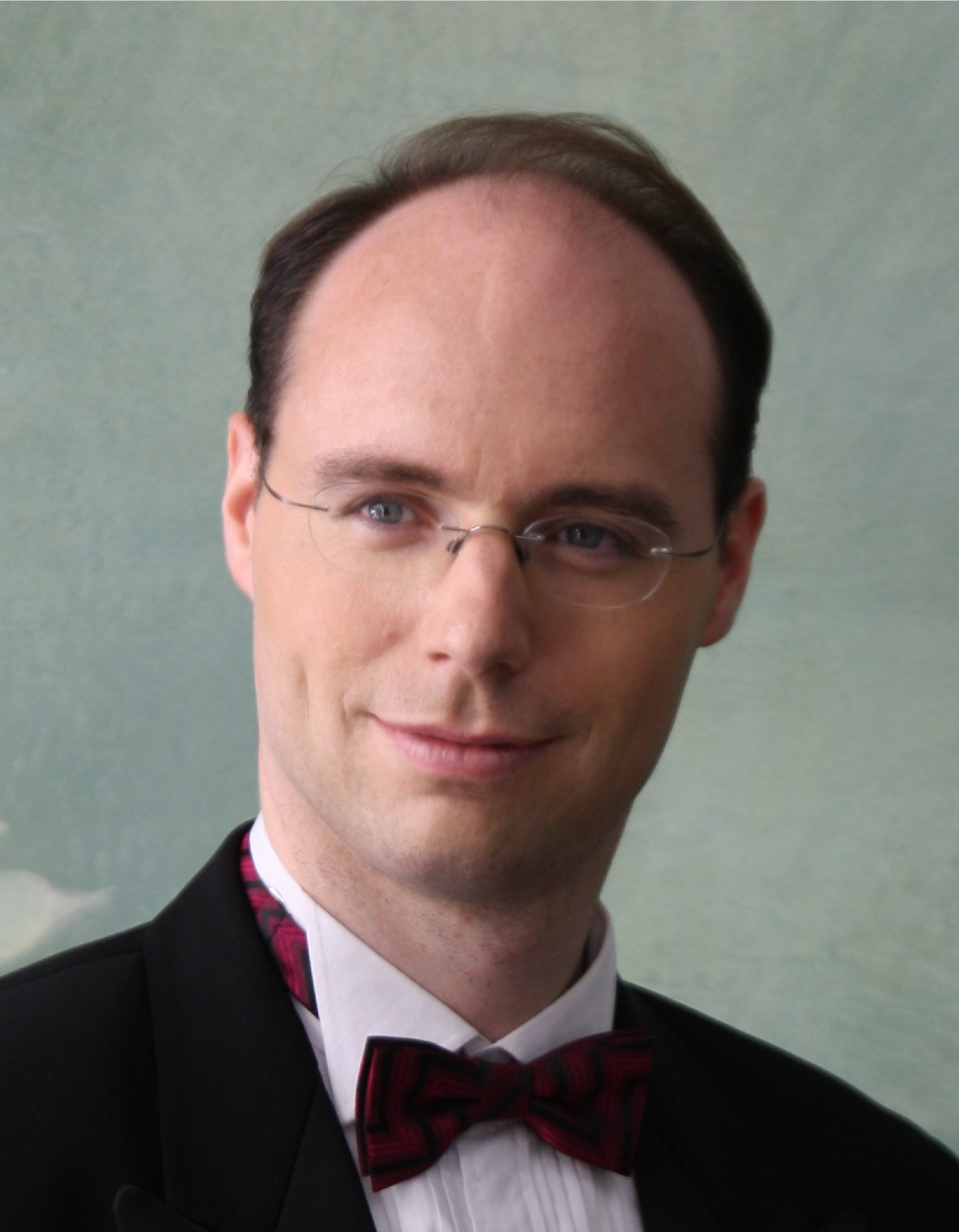}}]{Bruno Clerckx}
Bruno Clerckx is a (Full) Professor, the Head of the Wireless Communications and Signal Processing Lab, and the Deputy Head of the Communications and Signal Processing Group, within the Electrical and Electronic Engineering Department, Imperial College London, London, U.K. He received the M.S. and Ph.D. degrees in Electrical Engineering from the Université Catholique de Louvain, Louvain-la-Neuve, Belgium, in 2000 and 2005, respectively. From 2006 to 2011, he was with Samsung Electronics, Suwon, South Korea, where he actively contributed to 4G (3GPP LTE/LTE-A and IEEE 802.16m) and acted as the Rapporteur for the 3GPP Coordinated Multi-Point (CoMP) Study Item. Since 2011, he has been with Imperial College London, first as a Lecturer from 2011 to 2015, Senior Lecturer from 2015 to 2017, Reader from 2017 to 2020, and now as a Full Professor. From 2014 to 2016, he also was an Associate Professor with Korea University, Seoul, South Korea. He also held various long or short-term visiting research appointments at Stanford University, EURECOM, National University of Singapore, The University of Hong Kong, Princeton University, The University of Edinburgh, The University of New South Wales, and Tsinghua University. 
He has authored two books on “MIMO Wireless Communications” and “MIMO Wireless Networks”, 200 peer-reviewed international research papers, and 150 standards contributions, and is the inventor of 80 issued or pending patents among which 15 have been adopted in the specifications of 4G standards and are used by billions of devices worldwide. His research area is communication theory and signal processing for wireless networks. He has been a TPC member, a symposium chair, or a TPC chair of many symposia on communication theory, signal processing for communication and wireless communication for several leading international IEEE conferences. He was an Elected Member of the IEEE Signal Processing Society SPCOM Technical Committee. He served as an Editor for the IEEE TRANSACTIONS ON COMMUNICATIONS, the IEEE TRANSACTIONS ON WIRELESS COMMUNICATIONS, and the IEEE TRANSACTIONS ON SIGNAL PROCESSING. He has also been a (lead) guest editor for special issues of the EURASIP Journal on Wireless Communications and Networking, IEEE ACCESS, the IEEE JOURNAL ON SELECTED AREAS IN COMMUNICATIONS, the IEEE JOURNAL OF SELECTED TOPICS IN SIGNAL PROCESSING, and the PROCEEDINGS OF THE IEEE. He was an Editor for the 3GPP LTE-Advanced Standard Technical Report on CoMP. He is an IEEE Communications Society Distinguished Lecturer 2021-2022.
\end{IEEEbiography}




\end{document}